\documentclass[twocolumn]{aa}
\usepackage{graphicx}
\usepackage{natbib}
\bibpunct{(}{)}{;}{a}{}{,}
\begin{document}
   \title{Results of the ROTOR-program}

   \subtitle{II. The long-term photometric variability of weak-line T
     Tauri stars
   \thanks{$UBVR$ photometric data described in Table 1 are only
     available in electronic form at the CDS via anonymous ftp to
     cdsarc.u-strasbg.fr (130.79.128.5) or via http://cdsweb.u-
     strasbg.fr/cgi-bin/qcat?J/A+A/}}

   \author{K.N. Grankin \inst{1}, J. Bouvier \inst{2}, W. Herbst \inst{3}, \and S.Yu. Melnikov \inst{1}}

   \institute{Astronomical Institute of the Uzbek Academy of Sciences, Tashkent, Uzbekistan, 700052
           \and
              Laboratoire d'Astrophysique, Observatoire de Grenoble, Universit$\acute{\rm e}$ Joseph Fourier, B.P. 53, F-38041 Grenoble
Cedex 9, France
           \and
              Astronomy Department, Wesleyan University, Middletown, CT 06459
              }

\date{Received .../ Accepted ...}

\abstract
{T Tauri stars exhibit variability on all timescales,
  whose origin is still debated. On WTTS the variability is fairly
  simple and attributed to long-lived, ubiquitous cool spots.}
{We investigate the long term variability of WTTS, extending up to
  20 years in some cases, characterize it statistically and discuss
  its implications for our understanding of these stars.}
{We have obtained a unique, homogeneous database of photometric
  measurements for WTTS extending up to 20 years. It contains more
  than 9 000 UBV R observations of 48 WTTS. All the data were
  collected at Mount Maidanak Observatory (Uzbekistan) and they
  constitute the longest homogeneous record of accurate WTTS
  photometry ever assembled.}
{Definitive rotation periods for 35 of the 48 stars are obtained.
  Phased light curves over 5 to 20 seasons are now available for
  analysis. Light curve shapes, amplitudes and colour variations are
  obtained for this sample and various behaviors exhibited,
  discussed and interpreted.}
{Our main conclusion is that most WTTS have very stable long term
  variability with relatively small changes of amplitude or mean
  light level. The long term variability seen reflects modulation in
  the cold spot distributions. Photometric periods are stable over
  many years, and the phase of minimum light can be stable as well
  for several years. On the long term, spot properties do change in
  subtle ways, leading to secular variations in the shape and
  amplitudes of the light curves.}


\authorrunning{K.N. Grankin et al.}
\titlerunning{Results of the ROTOR-program. II. The long-term photometric variability of WTTS}

\maketitle
%

\section{Introduction}

The third catalog of pre-main-sequence emission-line stars
assembled by G. Herbig and collaborators includes 742 objects
(Herbig \& Bell 1988). Of these, about fifty objects were
classified as weak-line T Tauri stars (WTTS).These include objects
from X-ray surveys of several star-forming regions and from a
survey of stars with CaII H and K emis\-sion lines.WTTS exhibit a
weak, narrow $H\alpha$ line [W($H\alpha$) $<$ 10\AA], a strong LiI
absorption line (6707 \AA) with W(Li) $>$ 0.1 \AA, and a
substantial CaII H and K emission lines.

The first photoelectric UBVR observations of these stars revealed
periodic light variations in most WTTS (Rydgren \& Vrba 1983;
Rydgren et al. 1984; Bouvier et al. 1986, 1993, 1995; Grankin
1992, 1993, 1996; Vrba et al. 1993). Their photometric periods
ranged from 1 to 10 days and were in agreement with spectroscopic
estimates of their rotational velocities. It was assumed that the
periodic light variations were due to the nonuniform distribution
of cool photospheric spots over the stellar surface.

Rotation periods are now available for some hundreds of pre-main
sequence and recently arrived main sequence stars of solar-like
mass in five nearby young clusters: the Orion Nebula Cluster, NGC
2264, $\alpha$ Per, IC 2602 and the Pleiades (Mandel \& Herbst
1991; Attridge \& Herbst 1992; Eaton et al. 1995; Edwards et al.
1993; Choi \& Herbst 1996; Bouvier et al. 1997a). In combination
with estimates of stellar radii these data allow us to construct
distributions of surface angular momentum per unit mass at three
different epochs: nominally, 1, 2 and 50 Myr (see Herbst \& Mundt
2005). Recent work is extending the age range to 200 Myr and
adding many new clusters (Irwin et al. 2007).

The study of long-term variations in the main param\-eters of the
light curves for young spotted stars is also of considerable
interest. One would expect light-curve shapes to evolve as the
spots change size, shape, temperature, or location, and a
latitudinal migration of spots could cause changes in period, if
the surfaces are in differential rota- tion, as in the Sun. It
would, of course, be very interest- ing to find any cyclic pattern
that could be interpreted as evidence for a magnetic cycle.
Unfortunately, most of the photometric programs of observations of
WTTS stars have been carried out episodically, in an interval from
one to three months. Only one object (V410 Tau) has been
in\-vestigated in detail over several seasons (Vrba et al. 1988;
Herbst 1989; Petrov et al. 1994). These authors showed that there
were at least two extended and long-lived spot- ted regions on the
surface of V410 Tau during several years. The evolution of the
shape and amplitude of the light curve from season to season was
explained by changes in the spot sizes, temperatures, and relative
locations.

The first systematic photoelectric BVR observations for about
thirty WTTS in dark clouds in Taurus-Auriga began in 1990 at the
Maidanak Observatory (Grankin et al. 1995). During 1990-1993
rotation periods were dis- covered for 12 WTTS and refined for
another 9 WTTS. A fundamental result of the analysis of the
photometric behavior of these WTTS was the stability of the
initial epochs and rotation periods for 17 WTTS on a time scale
from 2 to 4 years. In addition, there is a high detection rate of
rotation periods among WTTS in the sample con- sidered. It is
generally believed that the stability of the initial epochs and
rotation periods over several years in- dicates that the active
region in each WTTS remains on a definite meridian over this time
scale. Variations in the maximum brightness levels, the amplitudes
and the shapes of the light curve for many WTTS is evidence for
migra- tion of the spots, within the limits of an active zone, and
for changes in their size. It is interesting to note, that RS CVn
and BY Dra stars show the same kind of long- term photometric
behaviour and that active longitudes have long been reported on
the Sun (Losh 1939). For RS CVn stars, active longitudes appear to
persist for a number of years (Butler 1996). Thus, an activity
mechanism with some similarities to that operating on the Sun may
be indicated by these common observational characteris- tics with
WTTS.

Longer term observations have shown, that the phe- nomenon of
stability of the initial epochs is most pro- nounced in seven
WTTS: LkCa 4, LkCa 7, V410 Tau, V819 Tau, V827 Tau, V830 Tau and
V836 Tau (Grankin 1997, 1998, 1999). Changes in the phase of
minimum light for these stars did not exceed $\pm 7\%$ of the
period on a time scale from 7 to 12 years. Similar results were
obtained for 23 WTTS in the young Orion Nebula Cluster (Choi \&
Herbst 1996).

Some interesting results have been received from anal\-ysis of
long-term light curves of 30 WTTS and 80 RS CVn and FK Com stars
(Grankin et al. 2002). These authors have found that in 18 of the
110 stars monitored, an in- crease in the amplitude of the
periodic variability accom- panies a rise of the maximum
brightness level. This result is difficult to understand in the
context of a simple model where the star contains only a single
spotted region in a high latitude zone that is always visible,
because increas- ing the size of the spot (as required to increase
the am- plitude) should decrease the maximum brightness of the
star. It is possible to explain such behavior by invoking a star
covered with mutiple spots at different latitudes. In this case
the amplitude will not depend just on the total area of the spots,
but will also depend on the distribu- tion of these spots on the
stellar surface. Similar results have been obtained recently for
24 WTTS stars in the ex- tremely young cluster IC 348 over five
observing seasons (Cohen et al. 2004). It has been shown that the
stability of average magnitude from season to season most likely
indi- cates that spots on these WTTS stars tend to redistribute
themselves and undergo changes in size, temperature, and location
rather than simply appear or disappear {\it en masse}.

The observations to date support to the possible exis- tence of
stellar activity cycles on WTTS. In an attempt to find and study
such activity cycles we have contin- ued to monitor some tens of
WTTS at Mt. Maidanak Observatory over many years. In this paper,
we employ the unique and extensive Mt. Maidanak database to in-
vestigate WTTS variability on a time scale of many years and, in
some cases, multiple decades. In Section 2, we de- scribe the
stellar sample and the observations carried out for the last 20
years at Mt. Maidanak. In Section 3, we dis- cuss the phenomenon
of rotational modulation of the light curves of 36WTTS. In Section
4, we describe a spot model and some results of modeling. In
Section 5, we define the statistical parameters used to
characterize the long term variability of WTTS. Further modeling
of the light curves will be presented in a later paper in this
series.

\section{Star sample and observations}

All $UBVR$ data were obtained at Mount Maidanak Observatory
(longitude: E$4^{\rm h}27^{\rm m}47^{\rm s}$; latitude:
$+38\degr41\arcmin$; altitude: 2709 m) in Uzbekistan. Results of
astroclimate studies at Maidanak observatory have been summarized
by Ehgamberdiev et al. (2000), who show that it is among the best
observatories in the world, in many respects. We targetted 48 WTTS
for long-term monitoring, mostly from the list of Herbig \& Bell
(1988). More than 9\,000 $UBVR$ magnitudes were collected for
these objects between 1990 and 2004, although the number of U
magnitudes is relatively small compared to the other colors due to
the faintness of the stars at those wavelengths. All the
observations were obtained at three telescopes (two 0.6 and one
0.48 m reflectors) using a single-channel pulse-counting
photometer with photomultiplier tubes. As a rule, each program
star was measured once per night at minimal airmass. We secured
between 12 and 30 $(U)BVR$ measurements for each sample target
during each observational season lasting several months. The
number of measurements depends on the object's visibility from Mt.
Maidanak. The largest number of measurements was thus obtained for
WTTS in Taurus, and fewer in Ophiuchus and Orion. Magnitudes are
provided in the Johnson $UBVR$ system. The rms error of a single
measurement in the instrumental system for a star brighter than
$12^m$ in $V$ is about $0^m.01$ in $BVR$ and $0^m.05$ in $U$.

Observations were carried out either differentially using a nearby
reference star (see Strauzis 1977) or directly by estimating the
nightly extinction (Nikonov 1976). In the latter case, several
reference stars were observed ev\-ery night to derive the
extinction coefficients in each fil\-ter. Selected standard stars
(Landolt 1983, 1992) were observed and used to transform
instrumental magnitudes into the Johnson/Cousins $UBVR$ system. We
then transformed the magnitudes to the Johnson $UBVR$ system using
the relationship from Landolt (1983)~: (V-R)$_C$ = -0.0320 +
0.71652$\times$(V-R)$_J$. The formal accuracy of this reduction
step is $0^m.01$. All observed local times have been converted to
heliocentric julian days (JDH). A detailed description of the
equipment, observation techniques, and data processing is given in
Shevchenko (1989).

Results of the Mt. Maidanak program of homoge- neous, long-term
photometry of WTTS are summarized in Table 1. Columns are: star's
name, star's number in the Herbig \& Bell (1988) catalogue,
spectral type, time span of observations in JD, number of seasons
observed, photometric range in the $V$ band, number of
observations in the V band, and average values of $B-V$ and $V-R$
colors. In the last column we give a multiplicity indica- tor, as
follows: spectroscopic binary (SB), visual binary (VB), and visual
tertiary (VT). The complete Maidanak database forWTTS is available
in electronic form at CDS, Strasbourg, via anonymous ftp to
cdsarc.u-strasbg.fr.

Most of the objects are located in the Taurus-Auriga region (40 of
the 48) and only 8 WTTS are in other star forming regions. The
range of spectral types is from G0 to M4. There are eight stars in
the range G0 to G9, eighteen between K0 and K5, ten between K6 and
K9, and nine between M0 and M4.

\section{Results of periodogram analysis}

The light curves obtained during the ROTOR campaign were analyzed
using the PERIOD software time-series analysis package (see
Starlink User Note). Periods were searched for each season using
two different methods, a chi-squared technique and the
Lomb-Scargle periodogram analysis. The first method is
straight-forward and the in- put data are simply folded on a
series of trial periods. At each trial period, the data are fitted
with a sine curve. The resulting reduced-$\chi^2$ values are
plotted as a function of trial frequency and the minima in the
plot suggest the most likely periods. See Horne et al. (1986) for
an example of the use of this method, which is ideally suited to
the study any sinusoidal variations. The second method is a novel
type of periodogram analysis, quite powerful for finding, and
testing the significance of, weak periodic signals in unevenly
sampled data (see Horne \& Baliunas 1986; Press \& Rybicki 1989).
As both methods have shown very similar results, here we simply
adopt the periods determined by the commonly used Lomb-Scargle
method and these are reported in Table 2.

False-alarm probabilities (FAP1) for the period were computed
using a Fisher randomization test, or Monte-Carlo method (see, for
example, Nemec \& Nemec 1985). FAP1 represents the proportion of
permutations (ie. shufied time-series) that contained a trough
lower than (in the case of the chi-squared method) or a peak
higher than (in the case of the Lomb-Scargle periodogram analysis)
that of the periodogram of the unrandomized dataset at any
frequency. This therefore represents the probability that, given
the frequency search parameters, no periodic component is present
in the data with this period. To ensure reliable signicance
values, the minimum number of permutations was set to 1000. If the
false alarm probabilities lie between 0.00 and 0.01, then the
quoted period is a correct one with 95\% confidence. Periods were
searched for within an interval ranging from 1 day to 20 days. The
periodogram is computed at 1000 frequencies between 0 and 0.05
d$^{-1}$.

A search for a periodic component has been made for all stars in
Table 1. Results of the periodogram analysis are listed in Table 2
for the 36 stars which showed a significant periodicity in one
season of observation or more. No significant periodicity was
found for the other 12 stars. We give the values of periods for
which the FAP1 lies between 0.00 and 0.01. If the false alarm
probabilities lies between 0.01 and 0.05, the period is designated
by a sign ":". If the FAP1 lies between 0.05 and 0.10, the period
is designated by a sign "?". Absence of any periodicity is
designated by a sign "-". In last column the most significant
period, based on a periodogram for the entire observing interval
(i.e. for all seasons) is presented.

We have tried to improve the period and ephemeris for each star in
Table 2 by the following procedure. Each seasonal lightcurve was
folded with its mean period to determine the time of minimum $T_0$
by fitting a polyno\-mial to the folded lightcurve. We used the
value for the period given in the last column of the Table 2  to
com\-pute the number of cycles elapsed between each of the
seasonal points $T_0$ and the time of minimum observed by us. Then
we computed the O-C diagram for the minima $T_0$ as a function of
cycle number. The residuals in this diagram can be minimized by
modifying the period. In this way a best fit period and new
ephemeris for each star are found. See Stelzer et al. (2003) for
an example of the use of this method for updating of the period
and ephemeris for V410 Tau. The new period and ephemeris for each
star are given in Table~3. We adopt this ephemeris and period for
all further analysis.

\section{Rotation modulation phenomena}

Thirty-six of our 48 WTTS exhibit periodic light variations during
one or more seasons. A list of these objects is given in Table 3.
Columns are: star's name, the epoch of our observations, the
minimum and maximum amplitudes of the periodic light variations
($\Delta V_{min}$ and $\Delta V_{max}$), the initial epoch, the
rotational period, and two columns of references. The first column
with references contains the papers in which there is a first
mention of a period. The second column with references specifies
the papers from which we have taken the initial epoch and the more
precise value for a period.

Minimum amplitudes of the periodic light variations are in the
range of $0^m.05$ to $0^m.4$ in the V band (see Figure~1) but more
than 80\% of stars have minimum amplitudes in the interval from
$0^m.05$ up to $0^m.1$. Maximum am\-plitudes of the periodic light
variations range from $0^m.15$ to $0^m.8$ in V and 75\% of the
stars show a maximum ampli\-tude within the limits of $0^m.15$ to
$0^m.3$. The peak of this distribution is at $0^m.15$. The stars
with the highest ampli- tudes of periodic light variations,
achieving $0^m.4-0^m.8$ in the V band, are: LkCa 4 ($0^m.79$),
V410 Tau ($0^m.63$), V836 Tau ($0^m.62$), LkCa 7 ($0^m.58$), V827
Tau ($0^m.51$), and V830 Tau ($0^m.45$). Such large amplitudes of
the light variation indicate the existence of very extended
spotted regions on the stellar surface. Note that some stars were
observed at fewer epochs than others, so we cannot be sure that
the quoted values of $\Delta V_{min}$ and $\Delta V_{max}$,
especially in these cases, are truly representative of the stars'
behavior on a time scale of decades. The actual extrema in these
values may be larger than is represented on Figure 1.

Of the 36 targets with periods, 35 lie within the range 0.48 to
9.9 days (see Table 3). The star TAP 4 shows the shortest period,
P= 0.482d, and TAP 45 shows the longest period of these 35, with
P= 9.9d. The most common periods are within the interval 0.5 to 4
d (see Figure 2). The exceptional star Is V501 Aur. Its
photometric period is close to 56 days, and it is most likely not
a rotation period. We may estimate the luminosity and radius of
this star, assuming, that it is associated with the Taurus dark
clouds at a distance of 140 pc. In that case, if the photometric
period corresponds to the period of rotation, its equatorial speed
of rotation would be about 3 km/s, which contradicts its measured
value or 25-27 km/s. It is possible that this object belongs to
another class of vari- able stars; spectral observations are
required to confirm or deny this guess.

Perhaps the most interesting result from our long-term
observations is that the phase light curve ($\varphi_{min}$) may
be conserved for times as long as several years. This phenomenon
of stability is most pronounced in V410 Tau (see Figure 3). Over
19 years of observation, the changes of $\varphi_{min}$ have not
exceeded 0.16 P, where P is the rotation period (Stelzer et al.
2003). We can see another example of phase stability extending for
15 years in LkCa~7 (Figure~4). Out of the total sample, 7 stars
show such stability of $\varphi_{min}$ for time intervals between
5 and 19 years, namely: LkCa 4, LkCa 7, V410 Tau, V819 Tau, V827
Tau, V830 Tau, and V836 Tau. Such long-term stability of
$\varphi_{min}$ is prob- ably due to the existence of so-called
active longitudes at which the (shorter-lived) extended spotted
regions are lo- cated (Grankin et al. 1995). Such active
longitudes have been reported on the Sun and RS CVn stars. It
should be noted, that the majority of these stars (6 of 7) also
are among the stars with the highest photometric amplitudes, i.e.
they are the most active variable stars in our sample. We suppose,
that the phenomenon of stability of $\varphi_{min}$ is connected
somehow with level of activity.

All the stars in our sample show significant changes in amplitude
and shape of phased light curves from season to season without any
dependence on their stability or in- stability of minimum phase.
For example, in the case of V410 Tau, the amplitude changes from
$0^m.39$ till $0^m.63$. In the case of LkCa~7 the amplitude
changes from $0^m.33$ to $0^m.58$. It is worth noting that the
most symmetric phased light curve around $\varphi_{min}$ generally
corresponds to the sea- son with the maximum amplitude. On the
other hand, the most asymmetrical phased light curves typically
cor- respond to seasons with the minimum amplitudes of light
variability (see Figure 3-4).We note that one object in our sample
showed a very unusual shape for a phased light curve in some
seasons. This is LkCa 4, which in 1992, 1993, and 1995 displayed a
flat portion near minimum in the phased light curve (see Figure
5). All other stars show constantly changing, in many cases nearly
sinusoidal light curves as a rule. In addition, this star has
shown the record amplitude in 2004, of about $0^m.8$ in the V
band. We do not know of another star of any type with such a large
amplitude caused by the phenomenon of rotational mod- ulation.

While some stars (such as V410 Tau, LkCa 7, and LkCa 4) show
gradual changes of amplitude and shape from season to season,
there are examples of other be- haviour. The amplitude of the
periodic light variations may appreciably change by as much as a
few tenths of a magnitude from one season to the next, as in case
of TAP 41. We can see from Figure 6 that the phased light curve
remains stable for two seasons, 1992 and 1993, after which its
initial phase and shape evolve rapidly. During such periods of
rapid light curve evolution, it is often not possible to detect
the star as a periodic variable or, if detected, its amplitude is
quite small, as in 1994 and 1998.

It may be noted, that for the majority of stars the aver- age
level of light does not change even as the amplitude de- creases
considerably. TAP 41 nicely illustrates this. It has been shown
that the stability of average level of light from season to season
most likely indicates that spots onWTTS tend to redistribute
themselves and undergo changes in size, temperature, and location
rather than to appear or disappear {\it en masse}. However, it is
also necessary to note, that some stars do show significant
changes in their av- erage level of light from season to season.
Examples are: TAP 50, V819 Tau, V827 Tau, V836 Tau, and VY Tau.

There is one more example of unusual photometric be- haviour among
our sample of WTTS. It may be seen on Figure 7 that the amplitude
of the light curve of V830 Tau changed by a few tenths of a
magnitude from its nor- mal value during the 2002 and 2003
seasons. During this period of time the phase light curve also had
a complex shape in the sense that two maxima and two minima were
observed per cycle. Such shapes of light curves can be the result
of the existence of two extended spotted regions on opposite sides
of the star. Similar light curves were ob- served at two epochs
for V410 Tau in 1981/1982 and in 1983/1984 (see Herbst 1989).

Preliminary analysis of our long-term observations re- vealed an
interesting relation between the variation of the amplitude of the
variability and the change in the level of maximum brightness
between different seasons. We no- ticed that some WTTS show an
increase of the amplitude of their variability accompanied with
the rise of the max- imum level of brightness. The maximum
brightness as a function of the amplitude of the light curve for 4
such WTTS are shown in Figure 8. The increase of the am- plitude
with the maximum level of brightness is clearly visible on this
Figure. This result is difficult to explain in a model of a star
spotted with a single, high latitude spot. Increasing the area of
such an almost-polar spot would de- crease the maximum brightness
while increasing the am- plitude of the light curve, which is
opposite to our mea- surements. A model with two nearly-polar
spots in oppo- site hemispheres can explain the observations if
the incli- nation angle of the star is sufficiently large so that
both spots contribute to the light variations. In this case, it
would be the motions of the spots in longitude that would account
for the amplitude variation, not the growth or decay of a spot
(see Grankin et al. 2002).

In this connection we have investigated how Vmin, Vmax and Vmean
change with amplitude. We found four groups of stars with
different dependences between Vmin, Vmax, Vmean and amplitude. The
first group of stars includes V410 Tau, LkCa 1, LkCa2, LkCa 7, TAP
26, TAP57, and SR 9. In all these cases, when the amplitude
increases, Vmax decreases (i.e. the star brightens), while Vmin
increases, and the mean light level remains about constant. In
Figure 9, the long term variations of Vmin, Vmax and Vmean as a
function of V amplitude for V410 Tau are presented. We can see
that Vmax increases, and Vmin decreases when the amplitude
increases. However, as the lower right panel of the figure shows,
there is no correlation between Vmax and Vmin.

A second group of stars exists which does show real dependence
between Vmax and Vmin (they decrease or increase simultaneously).
These stars are: LkCa 19, TAP 35, TAP 40, TAP 50, V819 Tau, V836
Tau, V1200 Tau and V1202 Tau. It may be noted, that these stars do
not show any reliable dependence between amplitude, Vmax and Vmin.
A third group of stars shows dependence be- tween amplitude and
Vmin, but does not show dependence between Vmax, Vmean and
amplitude. These are: Anon1, HD283572, LkCa 3, LkCa 4, Wa Oph/2,Wa
Oph/3, V827 Tau, and SR 12. A fourth group does not show any
depen- dences. These stars are: LkCa 11, V1197 Tau, V1207 Tau, Wa
Oph/1, VY Tau. In addition there are 3 stars which we cannot
classify unequivocally as belonging to any of these groups, namely
TAP 45, V830 Tau, V1199 Tau.

To summarize, all of the stars in Table 3 exhibit the phenomenon
of rotational modulation. But evolu- tion of phased light curves
occurs in a variety of man- ners. Some show specific dependencies
between amplitude, Vmin, Vmax and Vmean, but these are in various
direc- tions and some stars show no such dependencies. Detailed
modeling of these behaviors is beyond the scope of the present
paper and will be presented elsewhere. However, here we have used
a simple model of the spots to analyze some features of the light
curves and suggest reasons for the phenomena exhibited above.

\section{Preliminary Spot model and Some Results}

Models of light curves of TTs have been made based on assumed
geometries of the spots, e.g. one or two circular spots (Bouvier
\& Bertout 1989; Herbst 1989), while other models have used only
certain properties of the light vari- ations such as amplitude as
a function of wavelength to infer a spotted region's (SR) general
parameters, with- out assumptions about its geometry (Bouvier et
al. 1993). Assessment of advantages and disadvantages of
particular models is beyond the scope of our study and it is
certainly a good idea to model the data in a variety of ways. Here
we simply note that the non-uniqueness of light curve so- lutions
means that the choice of particular geometries for the spots needs
to be motivated by something other than the light curve data
themselves. For instance, Alekseev \& Gershberg (1996a)
justifiably criticize attempts to use parametric models to study
the surface nonuniformity of red dwarf stars with solar-type
activity. Such models nor- mally assume the existence of two
high-latitude spots near the star's magnetic poles, while
long-term photometric observations of red dwarfs provide evidence
for the exis- tence of low-latitude spotted zones. Accordingly,
Alekseev \& Gershberg (1996b) propose a model of zonal spottedness
for these stars, in which a collection of starspots is approx-
imated by two symmetric (about the equator) dark strips with a
variable spot-filling ratio in longitude. For WTTS, the situation
with the choice of a particular spottedness model is not yet so
certain.

On the one hand, the first detailed Doppler images of the surfaces
of the two brightest WTTS (V410 Tau and HDE 283572), which were
obtained by analyzing variations in the line profiles over the
rotation period, suggested that there were large cool spots at
high latitudes (Joncour et al. 1994a,b). The origin of such
high-latitude spots can be explained by the existence of a dipole
magnetic field, whose presence is assumed by some current models
that describe the interaction between a young star and a disk (see
for example Shu et al. 1994). On the other hand, we discussed
above that, based on the photometry of 36 periodic WTTS, there is
a similarity in behavior with main-sequence dwarfs, BY Dra stars,
and with RS CVn stars. So, it is arguable that some similarities
exist to solar-type activity. In this case, the long-term
evolution of the photometric light curves forWTTS may be improp-
erly explained in terms of a model with only one or two large,
high-latitude spots.

Here we will assume that there are many spots or groups of spots
on the surfaces of WTTS, including at high latitude. Analysis of
the Doppler images of the surface of V410 Tau obtained by Rice \&
Strassmeier (1996) provides evidence for the existence of both
high-latitude and low-latitude spots, consistent with this
assumption. We believe that the number of spots and their
distribution over the WTTS surface remains an open question.
Accordingly, for our quantitative analysis of the spottedness
parameters for the WTTS, we used a model whose parameters do not
depend on the number, shape, and positions of the spots. This
non-parametric model allows only the total area and mean
temperature of the spots in the visible stellar hemisphere to be
estimated. A detailed description of this model can be found in
Bouvier et al. (1993); Grankin (1998, 1999).

Vogt (1981) showed that the light variations of a star due to the
spottedness of its photosphere can be written as

$$
 \triangle m(\lambda)=-2.5\log\left\{\frac{1-[1-S(\lambda)]G_{max}(\lambda)}
     {1-[1-S(\lambda)]G_{min}(\lambda)}\right\},\eqno (1)
$$

\noindent where $\rm S(\lambda)$ is the ratio of the surface
brightness of the SR to the brightness of the photosphere; and
$\rm G_{max}(\lambda)$ and $\rm G_{min}(\lambda)$ are the spot
areas at maximum and minimum SR visibility, respectively. The
spot area depends on the wavelength via the limb-darkening
coefficient of the star. If we know the absolute maximum
brightness of the star, when its visible hemisphere is
completely free from spots ($\rm G_{min}=0$), then equation (1)
for the time of maximum SR visibility can be written as

$$
 m_{min}(\lambda) - m_{max}^{abs}(\lambda)
                     =-2.5\log\left\{1-[1-S(\lambda)]G_{max}(\lambda)
                       \right\},\eqno (2)
$$

\noindent where $\rm m_{max}^{abs}(\lambda)$ is the star's
absolute maximum brightness, and $\rm m_{min}(\lambda)$ is its
minimum brightness in a specific observing season.

If the limb-darkening effect is ignored, then the spot area can
be assumed to be independent of wavelength. In the case of a
broadband photometric system, relation (2) for V then takes the
form

$$
 V_{min} - V_{max}^{abs}
                     =-2.5\log\left\{1-[1-L_{V}]G_{max}
                       \right\},\eqno (3)
$$

\noindent where $\rm L_{V}=\left.(\int
I_{sp}(\lambda)\Phi_{V}(\lambda)d\lambda)\,\right/ (\int
I_{ph}(\lambda)\Phi_{V}(\lambda)d\lambda)$, \noindent $\rm
I_{ph}(\lambda)$ and $\rm I_{sp}(\lambda)$ - are the energy
distributions in the photosphere and the SR, respectively; $\rm
\Phi_{V}(\lambda)$ - is the response curve of the V filter; $\rm
V_{min}$ is the minimum V brightness for a specific observing
season; and $\rm V_{max}^{abs}$ is the absolute maximum V
brightness of the star. Equation (3) contains two free
parameters - the integral of the energy distribution for the SR,
which is uniquely related to the SR effective temperature, and
the spot area at maximum SR visibility. These two free
parameters can be determined by solving the system of two
equations of type (3) written for the V and R bands.

It may be noted that results of the computer simu- lation strongly
depend on the assumed spot-free bright- ness of the star. Many
authors assume that the observed maximum brightness of the star
($V_{max}$) corresponds to a spot-free configuration. In this
case, however, the phased light curve may be expected to have a
plateau at maxi- mum light., something that is rarely or never
seen. Also, in this case a star's maximum brightness could not
depend on the amplitude of variations, as found for some stars. It
may be seen on Fig. 8 that a decrease in periodicity am- plitude
is accompanied by a decrease in the star's max- imum brightness in
these cases. This demonstrates that, for these stars, the brighter
stellar hemisphere is not com- pletely free from spots. The
stellar photospheres of all TTs most likely remain spotted to some
degree at all phases of axial rotation. Thus, the observed maximum
brightness of a star is not the absolute maximum brightness of a
spot- ted star. It is quite clear, therefore, that only lower
limits on the parameters of spotted regions can be determined by
using the observed maximum brightness of the star as
($V^{abs}_{max}$) (see also the discussion in Bouvier et al.
1993).

We attempted to obtain more realistic estimates of the SR
parameters by using a linear relation between the amplitudes of
the periodicity in V and R. At minimum spot visibility,
$V_{max}^{abs}$ is related to $V_{max}$ by

$$
 V_{max} - V_{max}^{abs}
                     =-2.5\log\left\{1-[1-L_{V}]G_{min}
                       \right\}.\eqno (4)
$$

We see from (4) that for $V_{max}^{abs}$ to be determined, we
must know the degree of spottedness of the stellar surface at
minimum spot visibility. The following system of equations can
be used to estimate $G_{min}$:

$$ \left\{
\begin{array}{ccl}
V_{A}&=&-2.5\log\left(\frac{\displaystyle
1-\,\left[1-L_{V}\right]G_{max}} {\displaystyle
1-\left[1-L_{V}\right]G_{min}}\right)\\[8pt]
R_{A}&=&-2.5\log\left(\frac{\displaystyle
1-\,\left[1-L_{R}\right]G_{max}} {\displaystyle
1-\left[1-L_{R}\right]G_{min}}\right),\\
\end{array}
\right.\eqno (5) $$

\noindent where $\rm V_{A}$ and $\rm R_{A}$ are the amplitudes of
the periodicity in V and R, respectively. We derived the equation
for  $\rm V_{A}$ by subtracting equation (4) from (3). The
equation for $\rm R_{A}$ was derived in a similar way.

From the set of $\rm G_{min}$ values obtained by solving (5), we
choose only those which satisfy the relation

$$
 \rm V_{max}^{abs} - \rm R_{max}^{abs} \cong (V-R)_o, \eqno (6)
$$

\noindent where $\rm (V-R)_{\rm o}$ is the normal color index of
the star that corresponds to its spectral type; $\rm
V_{max}^{abs}$ and $\rm R_{max}^{abs}$ are the maximum V and R
brightness of the star estimated by substituting $\rm G_{min}$
into equation (4) written for the V and R bands, respectively.

In other words, the star's $\rm V_{max}^{abs} - \rm
R_{max}^{abs}$ color index corrected for the reddening due to
the spottedness of the stellar surface must essentially match
its normal color index. In this case, we obtain an upper limit
on $\rm V_{max}^{abs}$, which obviously differs from the
observed value $\rm V_{max}$. The SR parameters determined from
the upper limit on $\rm V_{max}^{abs}$ are overestimated
compared to their values calculated by assuming that
$V_{max}^{abs}=V_{max}$. For this reason, we call the SR
parameters calculated by the above method the upper limits on
the spottedness parameters.

With the goal of testing of our model we have estimated the
spottedness parameters for VY Ari and compared our results with
the results that were obtained by Alekseev \& Gershberg (1996a,b)
by using their model of zonal spottedness. The testing shows that
the two models yield similar results, despite the significant
differences between the two algorithms. For example, the
difference in the estimated effective spot temperatures ($\Delta
T$) is 100 K, while the difference in the estimated fractional
spot areas ($\Delta G/G$) does not exceed 2.5-3.5\%. All these
facts suggest that, despite its simplicity, our model allows the
main spot parameters to be estimated with a sufficient accuracy
(see Grankin 1998).

For the stars with the highest light curve amplitudes, this model
shows that the spotted regions cover from 17 to 73\% of the
visible stellar hemisphere and that the mean temperature of the
spots is 500-1400 K lower than the ambient photosphere. An
analysis of the calculated spot parameters of V410 Tau revealed
that there is a real correlation between the amplitude of
periodicity ($\Delta V$) and the spot distribution nonuniformity
($\Delta G$), which is defined as the difference of spot coverage
in the visible hemispheres of the star at maximum and minimum
light: $\Delta G = (G_{max} - G_{min})*100\%$, and $G_{max}$ and
$G_{min}$ are the area of the spotted region at maximum and
minimum spot visibility, respectively. As the amplitude increases
from $0^m.39$ to $0^m.63$, the degree of nonuniformity in the spot
distribution increases from 21 to 37\%. By contrast, the amplitude
is essentially independent of the variations in total spot area S,
where $S = 0.5*(G_{max} + G_{min})*100\%$. Therefore, the change
in the amplitude has to be caused by a change in the distribution
of the star spots rather than by a change in the total spot area.
This is shown in Fig. 10 on the upper panel, where we plot the
amplitude versus the degree of non-uniformity in the spot
distribution ($\Delta G$) and versus the total spot area (S).

Besides, the computer simulation has shown that the average
magnitude ($\overline{V}$) depends on the total spot area (S) and
it does not depend almost on the degree of non-uniformity in the
spot distribution ($\Delta G$). As the average magnitude decreases
from $10^m.80$ to $10^m.95$, the total spot area increases from 44
to 53\% (see Fig. 10, inferior panel). Thus, the cycles of star
activity should be exhibited in quasi-cyclic variations of the
average magnitude from season to season. Some WTTS from our sample
may demonstrate such quasi-cyclic variations of the average
magnitude. We are going to continue our photometric monitoring of
selected spotted stars in an effort to detect and study the cycles
of stellar-activity.

\section{The long term variability of WTTS}

In Paper I we have undertaken a statistical analysis of the
long-term light curves of classical T Tauri stars (CTTs) (Grankin
et al. 2007). We have presented and characterized the long term
photometric variations of 49 CTTs over as much as 20 years and
proposed an empirical classifica- tion scheme. We here apply the
same statistical analysis to the long term variability of WTTS.

As in Paper I, we first define a set of statistical param- eters
which allow us to characterize the long term photo- metric
variability of WTTS. In order to get robust sta- tistical
estimates, we consider only the 33 stars that have at least 5
observing seasons containing at least 12 pho- tometric
measurements per season. As shown in Paper I, this selection
provides reliable statistical estimates of their variability. For
each object, we compute the mean light level (${V_{m}}^{(i)}$) in
each season ~($i$) and the corresponding photometric range
($\Delta V^{(i)}$). The mean of (${V_{m}}^{(i)}$) over all seasons
yields the mean light level of the object ($\overline{V_m}$) while
its standard deviation ($\sigma_{V_m}$) measures the scatter of
the seasonal means. Similarly, we compute the average $V$ band
photometric amplitude over all seasons ($\overline{\Delta
V}=\frac{\sum \Delta V^{(i)}}{N_s}$, where $N_s$ -- number of
observational seasons) and its standard deviation ($\sigma_{\Delta
V}$). We also make use of the ratio $\frac{\sigma_{\Delta
V}}{\overline{\Delta V}}$. Two additional parameters were defined
in Paper~I, namely $C1 = <\frac{V_{med} - V_{min}}{\Delta V}>$,
where $V_{med}$ is the median light level of the object in a given
season, $V_{min}$ its maximum brightness level and $\Delta V$ its
photometric amplitude during this season and the bracket indicates
an average over all seasons, which measures the prefered
brightness state of the object, and $C2 =
\sigma_{V_{min}}/\sigma_{V_{max}}$, which characterizes the
relative seasonal variability of the maximum and minimum light
levels.

In addition to the statistical parameters above, we also
characterize the long term photometric behaviour of WTTS by
investigating their colour variations.  Previous studies of the
short term variability of WTTS have shown that most stars tend to
redden in colour indices (such as $V-R$) when fading (e.g. Herbst
et al. 1994)). This behavior can result from surface cold spots
(cf. Bouvier et al. 2005). Some illustrations of the linear
relations found for most objects between the $V-R$ colour and $V$
magnitude are given on Fig. 11. We characterize the colour
behaviour of WTTS by computing the color slopes
$\frac{\Delta(B-V)}{\Delta V}$ and $\frac{\Delta(V-R)}{\Delta V}$
from least-square fits, and the correlation coefficients
$\rho_{B-V}$ and $\rho_{V-R}$. We used the full set of data
pertaining to each object in order to compute these parameters.

The resulting values of the statistical and colour parameters for
33 WTTS whose light curves are shown in Fig.12a and 12b are listed
in Table~4. The columns in Table 4 provide the star's name, the
number of seasons $N_s^*$ with at least 12 measurements, the mean
brightness level ($\overline{V_m}$) and its seasonal standard
deviation ($\sigma_{V_m}$), the average photometric range
($\overline{\Delta V}$, averaged over all seasons) and its
seasonal standard deviation ($\sigma_{\Delta V}$), the fractional
variation of photometric amplitude as measured by the ratio
$\frac{\sigma_{\Delta V}}{\overline{\Delta V}}$, the color slope
$\frac{\Delta (B-V)}{\Delta V}$, their correlation coef- ficient
$\rho_{B-V}$, the color slope $\frac{\Delta (V-R)}{\Delta V}$
together with their correlation coefficient $\rho_{V-R}$, and the
parameters C1 and C2.

\subsection {General photometric properties}

WTTS usually exhibit a low level of variability. A histogram of
the average V-band amplitude for the 33 WTTS in our sample is
shown in Fig.~13. The distribution is quite peaked, with
$\overline{\Delta V}$ ranging from $0^m.1$ to $0^m.55$. Indeed, 28
WTTS (85\%) in our sample exhibit small ranges of variability
between $0^m.1$ and $0^m.3$, and only 5 (15\%) have average V-band
amplitudes larger than $0^m.4$ (LkCa~4, LkCa~7, SR~9, V410~Tau,
and V836~Tau). The mean light level of WTTS also appears very
stable over the years, with an exceedingly small standard
deviation $\sigma_{V_m}$. Its distribution is shown by the
histogram in Fig.~14, with the vast majority of WTTS having
$\sigma_{V_m}\leq 0^m.05$. Only one object, V836~Tau, exhibits
significant changes in average brightness over the years, with
$\sigma_{V_m}=0^m.2$. Similarly, the variations of photometric
amplitude from season to season are small, amounting to less than
$0^m.1$ for most WTTS, as shown by the histogram of
$\sigma_{\Delta V}$ in Fig.~15.

WTTS also exhibit a fairly straightforward $V-R$ colour behaviour,
turning redder when fainter. For most objects in our sample, the
$(V-R)$ index scales linearly with brightness, with a correlation
coefficient higher than 0.47 for 26 WTTS. An illustration is
provided in Fig.~11. Only 7 stars do not show any reliable
dependence of $V-R$ on brightness (correlation factor of lower
than $0.47$) when averaged over all seasons. These stars are:
SR~12, TAP~35, V501~Aur, V1197~Tau, V1207~Tau, Wa~Oph/2, and
Wa~Oph/3. The linear relation is replaced by a wide scatter of
points on the diagram for these stars.

A histogram of $\frac{\Delta{(V-R)}}{\Delta V}$ color slopes for
the 26 WTTS with linear color variations is shown in Fig.~16. It
strongly peaks at values of 0.2--0.4. In WTTS, brightness and
color variations result from the rotational modulation of the
stellar flux by cold surface spots. We showed in Paper I that CTTS
exhibit the same average $\frac{\Delta{(V-R)}}{\Delta V}$ color
slope as WTTS. This might be an indication that the photometric
variability of CTTS is also dominated by cold spots, but hot spots
as well as circumstellar extinction can produce this kind of $V-R$
color variations as well (Bouvier \& Bertout 1989).

A similar analysis was carried out for the $B-V$ index. Unlike the
$V-R$ index, $B-V$ variations show little correlation with
brightness, as seen from the low value of the correlation
coefficients listed in Table~4. Only 3 stars (TAP50, V410 Tau, and
V827 Tau) show a reliable dependence of $B-V$ on brightness
(correlation factor larger than $0.47$) when averaged over all
seasons. The lack of a correlation between brightness and  B-V
colour could be due to flares and/or chromospheric variability.

\subsection{Correlated properties}

In Paper I, we investigated the relationships between the various
statistical parameters we defined for CTTS. We do the same here for
WTTS and contrast the results with those obtained for CTTS.

Fig. 17 and 18 show plots of the statistical parameters against
each other that had allowed us in Paper I to identify various
photometric subgroups within the CTTS sample, corresponding to
specific sources of variability (cold spots, hot spots,
circumstellar extinction). In contrast, WTTS appear in all these
plots as an homogeneous pho- tometric group characterized by a low
level of variabil- ity as compared to CTTS. This confirms that the
pho- tometric variability of WTTS is dominated by a single
process, namely the rotational modulation of the stellar flux by
cold surface spots. In other words, the usually very stable long
term (years) photometric variability of WTTS merely reflects spot
modulation on much shorter timescales (weeks). Although we do not
find any clear cor- relation in WTTS between the various
variability proxies, the same trends found for CTTS over a much
larger pho- tometric range are seen, such as, e.g., a correlated
increase of average amplitude and its seasonal variations
(Fig.17).

The more complicated $C1$ and $C2$ parameters proved useful to
distinguish the various sources of photometric variability in CTTS
(see Paper I). WTTS are shown in the same plot in Fig.~18. Except
for one peculiar WTTS, V830 Tau, all WTTS scatter around the mean
value of these 2 parameters. This indicates that they have no
preferred brightness state ($C2$), and that their extreme
brightness levels ($C1$), high and low, vary by about the same
amount (see, e.g., LkCa7, V410 Tau, TAP 50 in Fig.~11b). The
isolation of V830 Tau on this diagram occurs because
$\sigma_{V_{max}}$ is three times greater than $\sigma_{V_{min}}$.
No WTTS are found in the lower left corner or in the upper right
corner of this plot, which were shown in Paper I to correspond to
UXor-type circumstellar extinction events and hot spot dominated
variability, respectively. This is again consistent with the cold
surface spots being the dominant and probably single source of
variability in WTTS.

Conversely, we note that many CTTS are found in the same location as
the WTTS in this diagram, even though CTTS photometric variability is
probably due to a mixture of hot and cold surface spots (see Paper
I). This is not unexpected, however~: as long as the surface spots
never disappear from view as the star rotates, the system's mean light
level as well as its extrema will be driven by spot modulation,
regardless of the spot's temperature. Hence, the system will have no
preferred brightness state ($C2\simeq$0.5) and its extreme values will
exhibit about the same amount of variability ($C1\simeq$1.). More
extreme values of the $C1$ and $C2$ parameters are expected only in
the particular case of a low latitude hot spot located at the surface
of a highly inclined system, in which case the spot disappears for
part of the rotational cycle.

That CTTS share the same location as WTTS in this diagram thus
does not imply that their source of variability is the same. It
does imply, however, that surface spots in most CTTS, as in most
WTTS, never disappear from view as the system rotates. In turn,
this suggests that hot accretion spots in CTTS are located at
relatively high latitudes on the stellar surface, as expected from
accretion from the inner disk edge onto a slightly tilted dipolar
magnetosphere. In addition, the impact of hot spots on photometric
variations is expected to be more conspicuous at shorter
wavelengths (e.g., U and B-bands) than at V (Bouvier \& Bertout
1989).

\section{Conclusions}

Our main conclusion is that most WTTS have very sta- ble long term
variability with relatively small changes of amplitude or mean
light level. The long term variability seen merely reflects
modulation in the cold spot distribu-tions. Photometric periods
are stable over many years, as is the phase of minimum light for
most objects, within some range. On the long term, the spot
properties do change in subtle ways, leading to some variations in
the shape and amplitudes of the light curves. The long term
variability of WTTS is thus perhaps best understood by assuming
that they are covered by a number of spots un- evenly distributed
on the stellar surface and located at preferential longitudes
and/or latitudes, much like active belts in the Sun and RS CVn
systems, rather than result- ing from a single polar spot. Such an
uneven spot distri- bution would account for the reported
correlation between amplitude and brightness level, as well as for
the quasi- sinusoidal shape of the light curves, which indicates
that the unspotted photosphere is not seen in these stars at any
rotational phase. It also implies that the coverage of the stellar
photosphere by cold spots may be much larger than anticipated from
simple modelling of the light curves, which only provide lower
limits to the spot areal coverage. Doppler imaging of a sample of
typical WTTS could be an effective means to obtain further
constraints on the distribution of cold spots at the surface of
WTTS.

Modelling of the light curves for V819 Tau, V827 Tau, V830 Tau and
V836 Tau indicates that the spotted regions cover as much as 30 to
90\% of the visible stellar hemisphere and that the mean
temperature of the spots is 500--1400~K lower than the ambient
photosphere. More detailed model calculations of the light curve
of V410 Tau showed that:

(i) The decrease in mean brightness is attributable to the increase
in total spot area from 47 to 53\% and that it is essentially
independent of the degree of nonuniformity in the spot distribution
over the surface.

(ii) The amplitude of the phased light curve depends on the degree
of nonuniformity in the spot distribution more strongly than on
the star's total spot area. An in- crease in amplitude was
accompanied by an increase in the degree of nonuniformity in the
spot distribution over the stellar surface from 21 to 35\%.

(iii) There is a weak correlation between the star's to- tal spot
area and its degree of nonuniformity in the spot distribution with
the correlation coefficient k=0.71. The increase in the star's
total spot area from 47 to 53\% was accompanied by a decrease in
the degree of nonuniformity in the spot distribution from 35 to
21\%.

The most active variable stars in our sample (i.e. those with the
largest amplitudes) demonstrate stability of phase of minimum
light most strongly over many years. The data presented here
should prove useful in fur- ther investigations of the nature and
evolution of spots on WTTS.

\begin{acknowledgements}

The authors wish to thank O. Ezhkova, M. Ibrahimov, V. Kondratiev,
S. Yakubov, S. Artemenko and many other observers and students,
who worked or are still working in the Tashkent Astronomical
Institute, for their participation in the photometric
observations. We are particularly grateful to C. Dougados, C.
Bertout, and F. M\'enard for their comments and suggestions
pertaining to this work. Support of the American Astronomical
Society, European Southern Observatory (grant A-02-048), and
International Science Foundation by Soros (grant MZA000) and a
CRDF grant (ZP1-341) is acknowledged. This work was also sup-
ported by a NATO Collaborative Linkage grant for European
countries (PST.CLG.976194) and by an ECO-NET program of the French
Ministry of Foreign Affairs, which we both grate- fully
acknowledge. W.H. gratefully acknowledges the support of NASA
through its Origins of Solar Systems Program. We especially thank
the referee F. Vrba for helpful comments on the original version
of this manuscript.

\end{acknowledgements}

\section{References}

\noindent Alekseev, I. Y. \& Gershberg, R. E. 1996a, AZh, 73, 589

\noindent Alekseev, I. Y. \& Gershberg, R. E. 1996b, AZh, 73, 579

\hangindent=0.3cm \noindent Attridge, J. M. \& Herbst, W. 1992,
Astrophys.J. (Lett.), 398, L61

\hangindent=0.3cm \noindent Berdnikov, L. N., Grankin, K. N.,
Chernyshev, A. V., et al. 1991, Soviet Astron. Lett., 117, 23

\noindent Bouvier, J. \& Bertout, C. 1989, A\&A, 211, 99

\hangindent=0.3cm \noindent Bouvier, J., Bertout, C., Benz, W., et
al. 1986, A\&A, 165, 110

\hangindent=0.3cm \noindent Bouvier, J., Boutelier, T., Alencar,
S., et al. 2005, in Protostars and Planets V (LPI Contribution No.
1286.), 8150

\hangindent=0.3cm \noindent Bouvier, J., Cabrit, S., Fernandez,
M., et al. 1993, A\&A, 272, 176

\hangindent=0.3cm \noindent Bouvier, J., Covino, E., Kovo, O., et
al. 1995, A\&A, 299, 89

\hangindent=0.3cm \noindent Bouvier, J., Forestini, M., \& Allain,
S. 1997a, A\&A, 326, 1023

\hangindent=0.3cm \noindent Bouvier, J., Wichmann, R., Grankin,
K., et al. 1997b, A\&A, 318, 495

\hangindent=0.3cm \noindent Butler, C. J. 1996, Proc. IAU
Symposium, 176, 423

\noindent Choi, P. I. \& Herbst, W. 1996, Astron.J., 111, 283

\hangindent=0.3cm \noindent Cohen, R., Herbst, W., \& Williams, E.
2004, Astron.J., 127, 1602

\hangindent=0.3cm \noindent Covino, E., Terranegra, L., Franchini,
M., et al. 1992, A\&AS, 94, 273

\hangindent=0.3cm \noindent Eaton, N. L., Herbst, W., \&
Hillenbrand, L. 1995, Astron.J., 110, 1735

\hangindent=0.3cm \noindent Edwards, S. E., Strom, S., \&
Hartigan, P. H. 1993, Astron.J., 106, 372

\hangindent=0.3cm \noindent Ehgamberdiev, S. A., Baijumanov, A.
K., Ilyasov, S. P., et al. 2000, A\&AS, 145, 293

\hangindent=0.3cm \noindent Ghez, A. M., Neugebauer, G., \&
Matthews, K. 1993, AJ, 106, 2005

\hangindent=0.3cm \noindent Grankin, K., Melnikov, S., Bouvier,
J., et al. 2007, A\&A, 461, 183

\noindent Grankin, K. N. 1991, Inform. Bull. Var. Stars., N3658

\noindent Grankin, K. N. 1992, Inform. Bull. Var. Stars., N3720

\noindent Grankin, K. N. 1993, Inform. Bull. Var. Stars., N3823

\noindent Grankin, K. N. 1994, Inform. Bull. Var. Stars., N4042

\noindent Grankin, K. N. 1996, Inform. Bull. Var. Stars., N4316

\noindent Grankin, K. N. 1997, Astron. Lett., 23, 700

\hangindent=0.3cm \noindent Grankin, K. N. 1997, in Low Mass Star
Formation - from Infall to Outflow., ed. F. Malbet \& A. Castets
(Proc. IAU Symposium N. 182. Dordrecht: D. Reidel Publ. Company.),
281

\noindent Grankin, K. N. 1998, Astron. Lett., 24, 580

\noindent Grankin, K. N. 1999, Astron. Lett., 25, 611

\hangindent=0.3cm \noindent Grankin, K. N., Granzer, T., \&
Strassmeier, K. 2002, in 1st Potsdam Thinkshop Poster Proceedings.
(Potsdam, Germany, May 6-10), 77

\hangindent=0.3cm \noindent Grankin, K. N., Ibragimov, M. A.,
Kondratiev, V. B., et al. 1995, AZh, 72, 894

\noindent Herbig, G. H. \& Bell, K. R. 1988, Lick Obs. Bull.,
1111, 1

\noindent Herbst, W. 1989, Astron.J., 98, 2268

\hangindent=0.3cm \noindent Herbst, W., Herbst, D. K., Grossman,
E. J., et al. 1994, AJ, 108, 1906

\noindent Herbst, W. \& Mundt, R. 2005, ApJ, 633, 967

\noindent Horne, J. H. \& Baliunas, S. L. 1986, AJ, 302, 757

\hangindent=0.3cm \noindent Horne, K., Wade, R. A., \& Szkody, P.
1986, MNRAS, 219, 791

\hangindent=0.3cm \noindent Irwin, J., et al. 2007, MNRAS, 380,
541

\hangindent=0.3cm \noindent Joncour, I., Bertout, C., \& Bouvier,
J. 1994a, A\&A, 291, L19

\hangindent=0.3cm \noindent Joncour, I., Bertout, C., \& Menard,
F. 1994b, A\&A, 285, L25

\hangindent=0.3cm \noindent Landolt, A. U. 1983, AJ, 88, 439

\hangindent=0.3cm \noindent Landolt, A. U. 1992, AJ, 104, 340

\hangindent=0.3cm \noindent Leinert, C., Zinnecker, H., Weitzel,
N., et al. 1993, A\&A, 278, 129

\hangindent=0.3cm \noindent Losh, H. M. 1939, Pub. Zurich Obs, 2

\hangindent=0.3cm \noindent Mandel, G. N. \& Herbst, W. 1991,
Astrophys.J. (Lett.), 383, L75

\hangindent=0.3cm \noindent Mathieu, R. D., Walter, F. M., \&
Myers, P. C. 1989, AJ, 98, 987

\hangindent=0.3cm \noindent Mundt, R., Walter, F. M., Feigelson,
E. D., et al. 1983, ApJ, 269, 229

\hangindent=0.3cm \noindent Nemec, A. F. L. \& Nemec, J. M. 1985,
AJ, 90, 2317

\hangindent=0.3cm \noindent Nikonov, V. 1976, Izvestiya KrAO, 54,
3

\hangindent=0.3cm \noindent Petrov, P., Shcherbakov, V.,
Berdyugina S. et al. 1994, Astron. Astrophys. Suppl. Ser., 107, 9

\hangindent=0.3cm \noindent Press, W. H. \& Rybicki, G. B. 1989,
ApJ, 338, 277

\hangindent=0.3cm \noindent Reipurth, B. \& Zinnecker, H. 1993,
A\&A, 278, 81

\hangindent=0.3cm \noindent Rice, J. \& Strassmeier, K. 1996,
A\&A, 316, 164

\hangindent=0.3cm \noindent Rydgren, A. E. \& Vrba, F. J. 1983,
ApJ, 267, 191

\hangindent=0.3cm \noindent Rydgren, A. E., Zak, D. S., Vrba, F.
J., et al. 1984, AJ, 89, 1015

\hangindent=0.3cm \noindent Shevchenko, V. S. 1989, Herbig Ae/Be
Stars (in Russian) (Tashkent: FAN)

\hangindent=0.3cm \noindent Shevchenko, V. S., Ezhkova, O. V.,
Kondratiev, V. B., et al. 1995, Inform. Bull. Var. Stars., N4206

\hangindent=0.3cm \noindent Shevchenko, V. S. \& Herbst, W. 1998,
AJ, 116, 1419

\hangindent=0.3cm \noindent Shu, F., Najita, J., Ruden, S., et al.
1994, ApJ, 429, 797

\hangindent=0.3cm \noindent Stelzer, B., Fernandez, M., Costa, V.
M., et al. 2003, A\&A, 411, 517

\hangindent=0.3cm \noindent Strauzis, V. 1977, Multicolor Stellar
Photometry (Vilnius: Mokslas)

\hangindent=0.3cm \noindent Vogt, S. S. 1981, ApJ, 250, 327

\hangindent=0.3cm \noindent Vrba, F. J., Chugainov, P. F., Weaver,
W. B., et al. 1993, AJ, 106, 1608

\hangindent=0.3cm \noindent Vrba, F. J., Herbst, W., \& Booth, J.
1988, Astron.J., 96, 1032

\hangindent=0.3cm \noindent Walter, F. M., Brown, A., Linsky, J.
L., et al. 1987, ApJ, 314, 297

\hangindent=0.3cm \noindent White, R. J. \& Ghez, A. M. 2001, ApJ,
556, 265

\hangindent=0.3cm \noindent Zakirov, M. M., Azimov, A. A., \&
Grankin, K. N. 1993, Inform. Bull. Var. Stars., N3898

\begin{table*}
\centering \caption[]{Results of Maidanak long-term photometry of
WTTS}

\begin{tabular}{lrllrcrllll}
\hline

Name      & HBC & SpT   &JD$_{min}$--JD$_{max}$ & $N_s$ & $V$
range & $N_{obs}$

& $\overline{B-V}$ & $\overline{V-R}$ & Multiplicity & Ref.\\

\hline

Anon 1    & 366 & M0    & 48953-50760 &  6 & 13.58-13.36 &  96 &
1.84 & 1.84 & & \\

HD 283572 & 380 & G6 IV & 48854-53300 & 13 &  9.16-8.89  & 425 & 0.77 & 0.70 & & \\

Hubble 4  & 374 & K7    & 49197-49251 &  1 & 12.71-12.59 &  27 & 1.62 & & &   \\

LkCa 1    & 365 & M4    & 49213-50760 &  5 & 13.80-13.64 &  88 & 1.45 & 1.71 & & \\

LkCa 2    &     & K7    & 48953-51157 &  7 & 12.37-12.19 & 119 & 1.39 & 1.34 & & \\

LkCa 3    & 368 & M1    & 48858-51157 &  7 & 12.17-11.97 & 144 & 1.48 & 1.52 & VB
(0."48) & wh01 \\

LkCa 4    & 370 & K7    & 48858-53289 & 13 & 13.19-12.28 & 284 & 1.42 & 1.37 & & \\

LkCa 5    & 371 & M2    & 48953-50081 &  4 & 13.63-13.44 &  65 & 1.47 & 1.50 & & \\

LkCa 7    & 379 & K7    & 48182-53300 & 15 & 12.75-12.11 & 422 & 1.36 & 1.35 & VB
(1."02) & wh01 \\

LkCa 11   &     & M2    & 48954-53291 & 13 & 13.33-13.10 & 197 & 1.50 & 1.58 & & \\

LkCa 14   & 417 & M0    & 48954-50081 &  4 & 11.84-11.63 &  63 & 1.19 & 1.05 & & \\

LkCa 16   & 420 & K7    & 48859-53289 &  4 & 12.71-12.32 &  61 & 1.50 & 1.41 & VB
(0."29) & wh01 \\

LkCa 19   & 426 & K0    & 48181-53300 & 13 & 11.08-10.80 & 308 & 1.01 & 1.09 & & \\

LkCa 21   & 382 & M3    & 48952-49251 &  2 & 13.54-13.34 &  32 & 1.57 & 1.65 & & \\

SR 9      & 264 & K5    & 46607-52827 & 15 & 11.86-11.28 & 336 & 1.25 & 1.21 & VB
(0."59) & gh93 \\

SR 12     & 263 & M1    & 46964-52819 & 10 & 13.53-13.09 & 195 & 1.57 & 1.60 & & \\

TAP 4     & 347 & K1    & 48858-49218 &  2 & 12.26-12.01 &  33 & 0.88 & 0.80 & & \\

TAP 9     & 351 & K5    & 48858-49635 &  3 & 12.25-12.08 &  82 & 1.05 & 0.99 & VB
(0."61) & le93 \\

TAP 10AB &352+353& G0+G5& 48953-49623 &  3 & 11.42-11.29 &  56 & 0.87 & 0.80 & VB
(8."6) & he88\\

TAP 11AB &354+355& K3+K0& 48853-49218 &  2 & 12.50-12.36 &  31 & 0.92 & 0.83 & & \\

TAP 14NE &356+357& K2   & 48853-49215 &  2 & 12.96-12.86 &  28 & 1.01 & 0.89 & & \\

TAP 26    & 376 & K7    & 48858-50760 &  6 & 12.44-12.16 & 165 & 1.13 & 1.01 & & \\

TAP 35    & 388 & K1    & 48860-53300 &  8 & 10.45-10.16 & 226 & 0.77 & 0.67 & & \\

TAP 40    & 392 & K5    & 48860-50760 &  6 & 12.69-12.52 & 137 & 1.17 & 1.05 & & \\

TAP 41    & 397 & K7    & 48860-53300 & 13 & 12.41-12.01 & 245 & 1.24 & 1.11 & & \\

TAP 45    & 403 & K7    & 48860-53291 &  8 & 13.36-13.11 & 145 & 1.45 & 1.33 & & \\

TAP 49    & 407 & G8    & 48860-50020 &  4 & 12.82-12.58 &  86 & 1.03 & 0.91 & & \\

TAP 50    & 408 & K0 IV & 49199-53311 & 12 & 10.34-10.06 & 263 & 0.89 & 0.79 & & \\

TAP 57NW  & 427 & K7    & 48857-53300 & 13 & 11.74-11.47 & 276 & 1.28 & 1.16 & SB
(P$\approx$ 2000d) & ma89\\

V501 Aur  &     & K2    & 49659-53311 & 10 & 10.83-10.54 & 247 & 1.62 & 1.35 & & \\

V410 Tau  &  29 & K3    & 46660-53300 & 19 & 11.24-10.52 & 959 & 1.16 & 1.04 & VT
(0."07, 0."29) & wh01 \\

V819 Tau  & 378 & K7    & 48126-53272 & 15 & 13.30-12.77 & 363 & 1.51 & 1.46 & VB
(10."5) & le93\\

V826 Tau  & 400 & K7    & 48135-49625 &  4 & 12.19-12.04 &  94 & 1.38 & 1.27 & SB
(P=3.9d) & mu83 \\

V827 Tau  & 399 & K7    & 48136-53299 & 14 & 13.00-12.24 & 236 & 1.42 & 1.37 & & \\

V830 Tau  & 405 & K7    & 48127-53298 & 15 & 12.39-11.91 & 383 & 1.35 & 1.24 & & \\

V836 Tau  & 429 & K7    & 48143-53292 & 11 & 14.05-13.01 & 170 & 1.53 & 1.44 & & \\

V1197 Tau &     & G8    & 49659-53300 & 10 & 10.66-10.49 & 232 & 0.84 & 0.75 & & \\

V1199 Tau &     & K0    & 49659-53311 & 10 & 10.75-10.53 & 208 & 0.72 & 0.66 & & \\

V1200 Tau &     & G5    & 49659-53311 & 10 & 11.35-11.15 & 212 & 0.93 & 0.81 & & \\

V1202 Tau &     & G0    & 49659-53311 & 10 & 10.95-10.68 & 208 & 0.83 & 0.71 & & \\

V1207 Tau &     & K7    & 49659-53300 & 10 & 12.01-11.81 & 174 & 1.24 & 1.08 & & \\

VY Tau    &  68 & M0    & 46383-52176 & 17 & 13.98-13.25 & 436 & 1.47 & 1.47 & VB
(0."66) & wh01 \\

Wa CrA/1  & 676 & K0 IV & 48049-48100 &  1 & 11.67-11.35 &  27 & 1.17 & 1.08 & & \\

Wa CrA/2  & 678 & G8 IV & 48049-48100 &  1 & 10.69-10.53 &  28 & 0.85 & 0.74 & & \\

Wa Oph/1  & 630 & K2 IV & 49145-53228 &  8 & 12.19-11.84 & 190 & 1.38 & 1.30 & & \\

Wa Oph/2  & 633 & K1 IV & 51733-53228 &  5 & 11.77-11.60 & 101 & 1.16 & 1.11 & & \\

Wa Oph/3  & 634 & K0 IV & 49145-53228 &  9 & 10.98-10.73 & 255 & 1.19 & 1.07 & & \\

Wa Oph/4  & 652 & K4    & 49147-52059 &  4 & 13.72-10.44 & 104 & 1.88 & 1.83 & VB
(8."7) & re93 \\

\hline

\multicolumn{11}{c}{ }\\

\multicolumn{11}{l}{Table 1. References}\\

\hline

\multicolumn{11}{l} {wh01: White \& Ghez (2001); re93: Reipurth \&
Zinnecker (1993); mu83: Mundt et al. (1983); ma89: Mathieu et al.
(1989);}\\

\multicolumn{11}{l} {le93: Leinert et al. (1993); he88: Herbig \&
Bell (1988); gh93: Ghez et al. (1993).}\\

\hline

\end{tabular}
\end{table*}

\newpage

\begin{table*}
\caption[]{The results of periodogram analysis.}\centering
{\tiny \tabcolsep=0.3em
\begin{tabular}{rllllllllllllllllllll}
\hline
      Name &  1986 &  1987 &  1988 &   1989 &  1990 &  1991 &  1992 &   1993 &   1994 &   1995 &   1996 &  1997 &  1998 &  1999 &   2000 &   2001 &   2002 &   2003 &   2004 & all data \\
\hline
    Anon 1 &       &       &       &        &       &       &       &   6.38 &   6.44 &   6.45 &      - &     - &       &       &        &        &        &        &        &      6.493 \\
 HD 283572 &       &       &       &        &       &       & 1.531 &  1.533 &  1.549 &  1.544 &  1.546 & 1.539 &       & 1.526 &  1.547 &  1.539 &  1.537 & 1.427? &  1.533 &      1.546 \\
    LkCa 1 &       &       &       &        &       &       &       & 2.502? & 2.494: &      - & 2.505: &     - &       &       &        &        &        &        &        &     2.513: \\
    LkCa 2 &       &       &       &        &       &       &       &      - &      - &  1.363 & 1.166? & 1.368 &       &       &        &        &        &        &        &      1.367 \\
    LkCa 3 &       &       &       &        &       &       &  7.30 &      - &   7.43 &   7.34 &   7.52 &  7.12 &       &       &        &        &        &        &        &       7.35 \\
    LkCa 4 &       &       &       &        &       &       & 3.376 &  3.367 &  3.373 &  3.376 &  3.376 & 3.373 & 3.381 & 3.373 &  3.376 &  3.376 &  3.376 &  3.376 &  3.387 &      3.376 \\
    LkCa 7 &       &       &       &        &  5.70 &  5.66 &  5.67 &   5.67 &   5.65 &   5.67 &   5.66 &  5.65 &  5.67 &  5.67 &   5.70 &   5.61 &   5.67 &   5.67 &   5.64 &      5.666 \\
   LkCa 11 &       &       &       &        &       &       &       &  1.542 & 1.541? &  1.540 &  1.543 &     - &     - & 1.539 &        &        &        &      - &        &     1.5396 \\
   LkCa 19 &       &       &       &        & 2.245 & 2.233 & 2.233 & 2.235? &      - &  2.237 &        &       & 2.227 & 2.240 &  2.237 &        &  2.240 &  2.238 &  2.238 &     2.228? \\
      SR 9 &  6.36 &  6.59 &  6.54 &      - &  6.54 & 6.68: &       &   6.53 &   6.53 &        &   6.56 &       & 6.29: &       &        &        &   6.34 &        &        &       6.55 \\
     SR 12 &       &       &     - & 3.407: & 3.497 & 3.578 &       &  3.506 &  3.410 &        &        &       &       &       &        &        &        &        &        &      3.515 \\
     TAP 4 &       &       &       &        &       &       & 0.482 &        &        &        &        &       &       &       &        &        &        &        &        &      0.482 \\
    TAP 26 &       &       &       &        &       &       & 0.714 &  0.714 & 0.714? &  0.833 &      - & 0.717 &       &       &        &        &        &        &        &    0.7134? \\
    TAP 35 &       &       &       &        &       &       &  2.78 &   2.74 &   2.75 &      - &      - &  2.88 &       &       &        &        &        &   2.75 &      - &          - \\
    TAP 40 &       &       &       &        &       &       & 1.553 &  1.552 &      - &      - &      - &     - &       &       &        &        &        &        &        &      1.555 \\
    TAP 41 &       &       &       &        &       &       & 2.424 &  2.421 &  2.417 &  2.421 &  2.427 & 2.417 & 2.421 & 2.424 &        &        &        &  2.418 &  2.429 &      2.427 \\
    TAP 45 &       &       &       &        &       &       &     - &   9.73 &      - &      - &  9.98? & 9.98: &       &       &        &        &        &  9.76: &        &       9.90 \\
    TAP 49 &       &       &       &        &       &       & 3.30? &  3.14? &      - &      - &        &       &       &       &        &        &        &        &        &          - \\
    TAP 50 &       &       &       &        &       &       &       &   3.04 &   3.01 &      - &   3.05 &  3.04 &       &  3.02 &        &      - &      - &      - &   3.03 &     3.017? \\
  TAP 57NW &       &       &       &        &       &       &  9.35 &   9.46 &      - &   9.37 &  9.22? &  9.37 & 9.15: &     - &        &        &   9.11 &  9.32: &   9.71 &       9.11 \\
  V501 Aur &       &       &       &        &       &       &       &        &     43 &     56 &     56 &    52 &    52 &     - &        &        &      - &      - &     61 &       55.6 \\
  V410 Tau & 1.871 & 1.873 & 1.873 &  1.872 & 1.873 & 1.872 & 1.872 &  1.872 &  1.872 &  1.872 &  1.872 & 1.873 & 1.874 & 1.872 &  1.877 &  1.872 &  1.872 &  1.872 &  1.874 &     1.8718 \\
  V819 Tau &       &       &       &        &  5.58 &  5.61 &  5.54 &   5.50 &   5.51 &   5.51 &   5.51 &  5.51 &       &  5.53 &   5.49 &   5.52 &  5.51: &   5.48 &   5.73 &     5.533: \\
  V827 Tau &       &       &       &        &  3.76 &  3.77 &  3.75 &   3.76 &   3.76 &      - &   3.76 &  3.76 &  3.77 &  3.70 &        &        &        &        &   3.77 &      3.759 \\
  V830 Tau &       &       &       &        & 2.738 & 2.744 & 2.738 &  2.740 &  2.740 &  2.742 &  2.740 & 2.738 & 2.745 & 2.740 & 2.745: & 2.677? &      - &      - &        &      2.742 \\
  V836 Tau &       &       &       &        &  6.75 & 6.68? &  6.77 &        &   6.75 &   6.77 &   6.76 &  6.80 &       &       &        &        &        &        &        &      6.757 \\
 V1197 Tau &       &       &       &        &       &       &       &        &   2.67 &      - &      - &  2.65 &  2.64 &  2.66 &        &      - &      - &      - &   2.67 &      2.628 \\
 V1199 Tau &       &       &       &        &       &       &       &        &  1.818 &      - &      - & 1.824 &     - & 1.815 &        &      - & 1.840? & 1.804: & 1.827? &     1.810: \\
 V1200 Tau &       &       &       &        &       &       &       &        &  1.599 &  1.592 &      - & 1.605 & 1.592 &     - &        &      - &      - &      - &  1.606 &      1.596 \\
 V1202 Tau &       &       &       &        &       &       &       &        &  2.712 & 2.663: &      - & 2.686 & 2.685 & 2.740 &        &        &      - &      - &  2.681 &      2.695 \\
 V1207 Tau &       &       &       &        &       &       &       &        &   7.74 &   7.80 &      - & 7.53: &     - &  7.53 &        &        &   7.72 &        &   7.75 &       8.10 \\
    VY Tau &       &  5.39 &  5.38 &   5.35 &     - &     - &     - &   5.30 &        &   5.35 &  5.48: &  5.44 &     - &     - &      - &        &        &        &        &      5.369 \\
  Wa CrA/1 &       &       &       &        &  2.25 &       &       &        &        &        &        &       &       &       &        &        &        &        &        &       2.25 \\
  Wa CrA/2 &       &       &       &        &  2.80 &       &       &        &        &        &        &       &       &       &        &        &        &        &        &       2.80 \\
  Wa Oph/1 &       &       &       &        &       &       &       &   3.82 &   3.81 &        &        &       &       &       &   3.78 &   3.82 &   3.80 &   3.77 &  3.78? &      3.71: \\
  Wa Oph/3 &       &       &       &        &       &       &       &  1.523 &  1.517 &        &  1.526 &       &       &       &  1.527 &  1.455 &  1.501 &      - &  1.518 &      1.527 \\
\hline
\end{tabular}
}
\end{table*}

\newpage

\begin{table*}
\caption[]{WTTS with well-known periods.}\centering
\begin{tabular}{lcclclcc}
\hline
Name & Epoch of & $\Delta V_{min}$ & $\Delta V_{max}$ & Initial Epoch & Period &
\multicolumn{2}{c}{References}\\
     &   observ.  & (mag)            & (mag)            & JDH 2400000+  & (day)  & (1) & (2) \\
\hline
         Anon 1 &  1992-1997 & 0.11 & 0.20 & 48951.7 &  6.493   & gr96 & gr97 \\
      HD 283572 &  1992-2004 & 0.08 & 0.24 & 48850.3 &  1.529   & wa87 & gr97 \\
         LkCa 1 &  1993-1997 & 0.06 & 0.14 & 49212.8 &  2.497   & gr96 & gr97 \\
         LkCa 2 &  1992-1998 & 0.05 & 0.18 & 49950.3 &  1.3614  & gr96 & this paper \\
         LkCa 3 &  1992-1998 & 0.08 & 0.16 & 48856.5 &  7.35    & gr96 & this paper \\
         LkCa 4 &  1992-2004 & 0.35 & 0.79 & 46296.8 &  3.374   & gr93 & this paper \\
         LkCa 7 &  1990-2004 & 0.33 & 0.58 & 45998.9 &  5.6638  & gr92 & gr97 \\
        LkCa 11 &  1993-2004 & 0.12 & 0.19 & 48952.0 &  1.5396  & gr96 & this paper \\
        LkCa 19 &  1990-2004 & 0.09 & 0.15 & 48179.8 &  2.236   & gr93 & gr97 \\
           SR 9 &  1986-2003 & 0.24 & 0.55 & 46607.0 &  6.531   & bo89 & this paper \\
          SR 12 &  1987-2003 & 0.15 & 0.41 & 46963.5 &  3.516   & bo89 & this paper \\
          TAP 4 &  1992-1993 & 0.08 & 0.22 & 48853.3 &  0.482   & this paper & this paper \\
         TAP 26 &  1992-1997 & 0.08 & 0.27 & 48619.6 &  0.7135  & gr93 & gr97 \\
         TAP 35 &  1992-2004 & 0.05 & 0.13 & 48858.5 &  2.734   & gr93 & gr97 \\
         TAP 40 &  1992-1997 & 0.09 & 0.13 & 48859.8 &  1.5548  & gr93 & gr97 \\
         TAP 41 &  1992-2004 & 0.11 & 0.39 & 48857.9 &  2.425   & gr93 & this paper \\
         TAP 45 &  1992-2004 & 0.09 & 0.21 & 48859.5 &  9.909   & gr94 & gr97 \\
         TAP 49 &  1992-1995 & 0.07 & 0.12 & 48857.2 &  3.32    & gr94 & gr97 \\
         TAP 50 &  1993-2004 & 0.05 & 0.13 & 49195.3 &  3.039   & gr94 & this paper \\
       TAP 57NW &  1992-2004 & 0.10 & 0.21 & 48848.7 &  9.345   & za93 & gr97 \\
V501 Aur  (W185)&  1994-2004 & 0.05 & 0.24 & 49625.7 & 55.95    & bo97 & this paper \\
       V410 Tau &  1986-2004 & 0.39 & 0.63 & 46659.6 &  1.87197 & ry83 & st03 \\
       V819 Tau &  1990-2004 & 0.12 & 0.28 & 48122.4 &  5.53113 & ry84 & this paper \\
       V827 Tau &  1990-2004 & 0.07 & 0.51 & 48136.3 &  3.75837 & bo86 & this paper \\
       V830 Tau &  1990-2004 & 0.15 & 0.45 & 48124.7 &  2.74101 & ry84 & this paper \\
       V836 Tau &  1990-2004 & 0.39 & 0.62 & 48142.9 &  6.75791 & ry84 & this paper \\
V1197 Tau (W123)&  1994-2004 & 0.06 & 0.16 & 49658.0 &  2.662   & bo97 & this paper \\
V1199 Tau (W130)&  1994-2004 & 0.06 & 0.21 & 49658.0 &  1.8122  & bo97 & this paper\\
V1200 Tau (W139)&  1994-2004 & 0.08 & 0.13 & 49659.3 &  1.605   & bo97 & this paper \\
V1202 Tau (W145)&  1994-2004 & 0.06 & 0.19 & 49658.8 &  2.7136  & bo97 & this paper\\
V1207 Tau (W188)&  1994-2004 & 0.06 & 0.14 & 49656.9 &  7.741   & bo97 & this paper\\
         VY Tau &  1985-2001 & 0.11 & 0.31 & 44610.7 &  5.36995 & gr91 & gr94 \\
       Wa CrA/1 &  1990      & 0.32 & 0.32 & 48048.3 &  2.24    & sh95 & sh95 \\
       Wa CrA/2 &  1990      & 0.16 & 0.16 & 48048.3 &  2.79    & co92 & sh95 \\
       Wa Oph/1 &  1993-2004 & 0.10 & 0.30 & 48986.1 &  3.792   & za93 & this paper \\
       Wa Oph/3 &  1993-2004 & 0.06 & 0.23 & 49144.3 &  1.5214  & sh98 & this paper \\
\hline

\multicolumn{8}{c}{ }\\

\multicolumn{8}{l}{Table 3. References}\\

\hline

\multicolumn{8}{l}{be91: Berdnikov et al. (1991); bo86: Bouvier et
al. (1986); bo89: Bouvier \& Bertout (1989); bo97: Bouvier et al.
(1997b);}\\

\multicolumn{8}{l}{co92: Covino et al. (1992); gr91: Grankin
(1991); gr92: Grankin (1992); gr93: Grankin (1993); gr94: Grankin
(1994);}\\

\multicolumn{8}{l}{gr96: Grankin (1996); gr97: Grankin (1997);
ry83: Rydgren \& Vrba (1983); ry84: Rydgren et al. (1984); sh95:
}\\

\multicolumn{8}{l}{Shevchenko et al. (1995); sh98: Shevchenko \&
Herbst (1998); st03: Stelzer et al. (2003); wa87: Walter et al.
(1987);}\\

\multicolumn{8}{l}{za93: Zakirov et al. (1993).}\\

\hline
\end{tabular}
\end{table*}

\begin{table*}
\caption[]{Statistical properties of WTTS light curves (see
text).}\centering
\begin{tabular}{lrrllllllllll}
\hline
Name &    $N_s^*$    &$\overline{V_m}$& $\sigma_{V_m}$ &$\overline{\Delta V}$ & $\sigma_{\Delta V}$
& $\frac{\sigma_{\Delta V}}{\overline{\Delta V}}$ & $\frac{\Delta(B-V)}{\Delta V}$ & $\rho_{B-V}$ &
$\frac{\Delta(V-R)}{\Delta V}$ & $\rho_{V-R}$ & $ c1 $ & $ c2 $\\
\hline
    Anon 1 &  5 & 13.445 & 0.015 & 0.156 & 0.036 & 0.230 &  0.142 &  0.13 & 0.376 & 0.60 & 0.503 & 0.175\\
HD 283572  & 12 &  9.000 & 0.028 & 0.133 & 0.045 & 0.342 &  0.245
&  0.44 & 0.246 & 0.56 & 0.486 & 0.551\\
    LkCa 1 &  5 & 13.710 & 0.007 & 0.097 & 0.033 & 0.343 & -0.363 & -0.22 & 0.709 & 0.72 & 0.407 & 0.841\\
    LkCa 2 &  5 & 12.286 & 0.016 & 0.096 & 0.053 & 0.549 & -0.001 &  0.00 & 0.338 & 0.68 & 0.592 & 1.034\\
    LkCa 3 &  6 & 12.034 & 0.017 & 0.118 & 0.029 & 0.245 & -0.041 & -0.07 & 0.290 & 0.59 & 0.341 & 0.437\\
    LkCa 4 & 13 & 12.624 & 0.078 & 0.512 & 0.122 & 0.238 &  0.035 &  0.22 & 0.188 & 0.79 & 0.489 & 0.432\\
    LkCa 7 & 15 & 12.423 & 0.025 & 0.438 & 0.078 & 0.178 &  0.112 &  0.39 & 0.228 & 0.81 & 0.481 & 0.924\\
   LkCa 11 & 10 & 13.213 & 0.022 & 0.154 & 0.025 & 0.164 &  0.004 &  0.01 & 0.302 & 0.48 & 0.457 & 1.249\\
   LkCa 19 & 12 & 10.915 & 0.041 & 0.117 & 0.023 & 0.200 &  0.092 &  0.19 & 0.309 & 0.69 & 0.540 & 0.848\\
      SR 9 & 12 & 11.524 & 0.048 & 0.417 & 0.099 & 0.239 &  0.152 &  0.37 & 0.208 & 0.75 & 0.377 & 0.796\\
     SR 12 &  6 & 13.288 & 0.026 & 0.303 & 0.087 & 0.288 &  0.226 &  0.25 & 0.218 & 0.44 & 0.495 & 0.254\\
    TAP 26 &  6 & 12.320 & 0.044 & 0.164 & 0.064 & 0.390 &  0.045 &  0.12 & 0.368 & 0.72 & 0.559 & 1.380\\
    TAP 35 &  8 & 10.276 & 0.063 & 0.089 & 0.031 & 0.351 &  0.087 &  0.32 & 0.082 & 0.26 & 0.477 & 0.822\\
    TAP 40 &  6 & 12.608 & 0.022 & 0.109 & 0.018 & 0.160 & -0.145 & -0.19 & 0.286 & 0.47 & 0.484 & 1.474\\
    TAP 41 & 12 & 12.181 & 0.025 & 0.236 & 0.090 & 0.380 &  0.082 &  0.24 & 0.170 & 0.53 & 0.543 & 0.666\\
    TAP 45 &  8 & 13.205 & 0.031 & 0.131 & 0.038 & 0.292 & -0.086 & -0.11 & 0.259 & 0.47 & 0.473 & 0.770\\
    TAP 50 & 10 & 10.197 & 0.044 & 0.088 & 0.027 & 0.309 &  0.320 &  0.61 & 0.324 & 0.67 & 0.488 & 0.994\\
  TAP 57NW & 13 & 11.620 & 0.025 & 0.153 & 0.044 & 0.290 &  0.162 &  0.23 & 0.359 & 0.59 & 0.514 & 0.961\\
  V501 Aur &  9 & 10.637 & 0.044 & 0.142 & 0.070 & 0.492 &  0.120 &  0.29 & 0.124 & 0.31 & 0.387 & 0.255\\
  V410 Tau & 19 & 10.863 & 0.044 & 0.534 & 0.073 & 0.137 &  0.113 &  0.58 & 0.154 & 0.79 & 0.406 & 0.978\\
  V819 Tau & 14 & 13.050 & 0.113 & 0.197 & 0.046 & 0.231 &  0.105 &  0.35 & 0.203 & 0.73 & 0.550 & 1.103\\
  V827 Tau & 12 & 12.528 & 0.105 & 0.252 & 0.127 & 0.505 &  0.140 &  0.48 & 0.234 & 0.82 & 0.475 & 0.684\\
  V830 Tau & 14 & 12.183 & 0.043 & 0.253 & 0.081 & 0.319 &  0.103 &  0.30 & 0.207 & 0.68 & 0.416 & 2.776\\
  V836 Tau &  8 & 13.382 & 0.196 & 0.500 & 0.084 & 0.168 &  0.020 &  0.07 & 0.201 & 0.86 & 0.422 & 0.973\\
 V1197 Tau & 10 & 10.564 & 0.019 & 0.095 & 0.030 & 0.312 &  0.203 &  0.25 & 0.290 & 0.37 & 0.477 & 0.821\\
 V1199 Tau & 10 & 10.643 & 0.031 & 0.099 & 0.046 & 0.464 &  0.041 &  0.08 & 0.218 & 0.50 & 0.476 & 1.044\\
 V1200 Tau &  9 & 11.247 & 0.028 & 0.098 & 0.017 & 0.179 &  0.137 &  0.22 & 0.299 & 0.50 & 0.552 & 0.893\\
 V1202 Tau & 10 & 10.811 & 0.036 & 0.112 & 0.035 & 0.313 &  0.204 &  0.40 & 0.292 & 0.58 & 0.472 & 0.887\\
 V1207 Tau &  9 & 11.920 & 0.026 & 0.092 & 0.028 & 0.308 &  0.062 &  0.09 & 0.250 & 0.46 & 0.501 & 1.057\\
     VYTau & 15 & 13.650 & 0.080 & 0.218 & 0.091 & 0.419 &  0.286 &  0.30 & 0.356 & 0.70 & 0.542 & 1.445\\
  Wa Oph/1 &  7 & 12.007 & 0.065 & 0.165 & 0.072 & 0.436 & -0.083 & -0.14 & 0.338 & 0.68 & 0.457 & 1.015\\
  Wa Oph/2 &  5 & 11.676 & 0.025 & 0.114 & 0.032 & 0.285 &  0.276 &  0.34 & 0.196 & 0.36 & 0.445 & 0.672\\
  Wa Oph/3 &  8 & 10.835 & 0.023 & 0.149 & 0.054 & 0.362 &  0.200 &  0.30 & 0.210 & 0.45 & 0.484 & 0.542\\
\hline
\end{tabular}
\end{table*}

\newpage

\begin{figure*}
\begin{minipage}[c]{8.7cm}
\includegraphics[width=8.5cm]{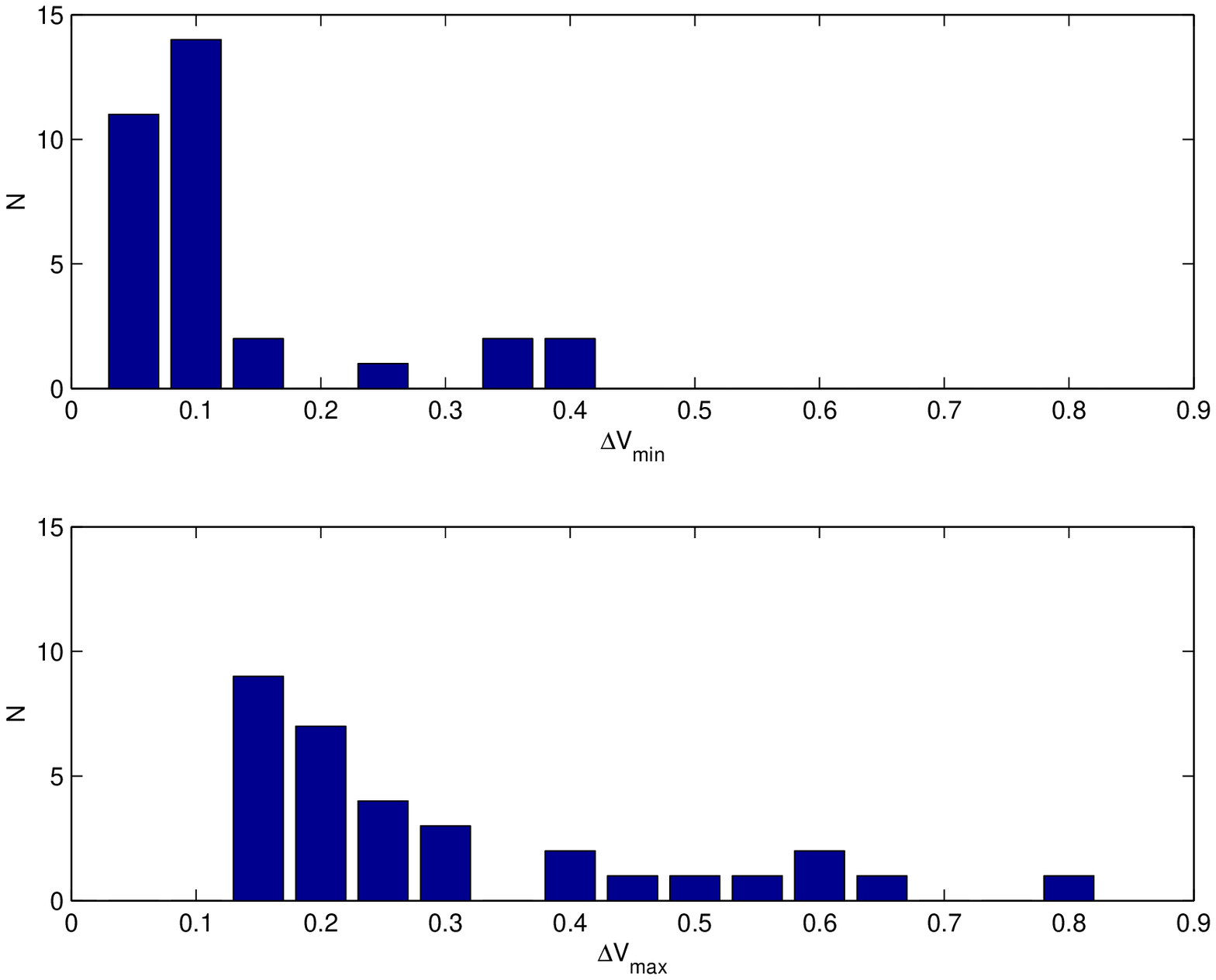}
Fig.1. Distribution of minimum and maximum amplitudes among WTTS.
\end{minipage}
\hfill
\begin{minipage}[c]{8.6cm}
\includegraphics[width=8.4cm]{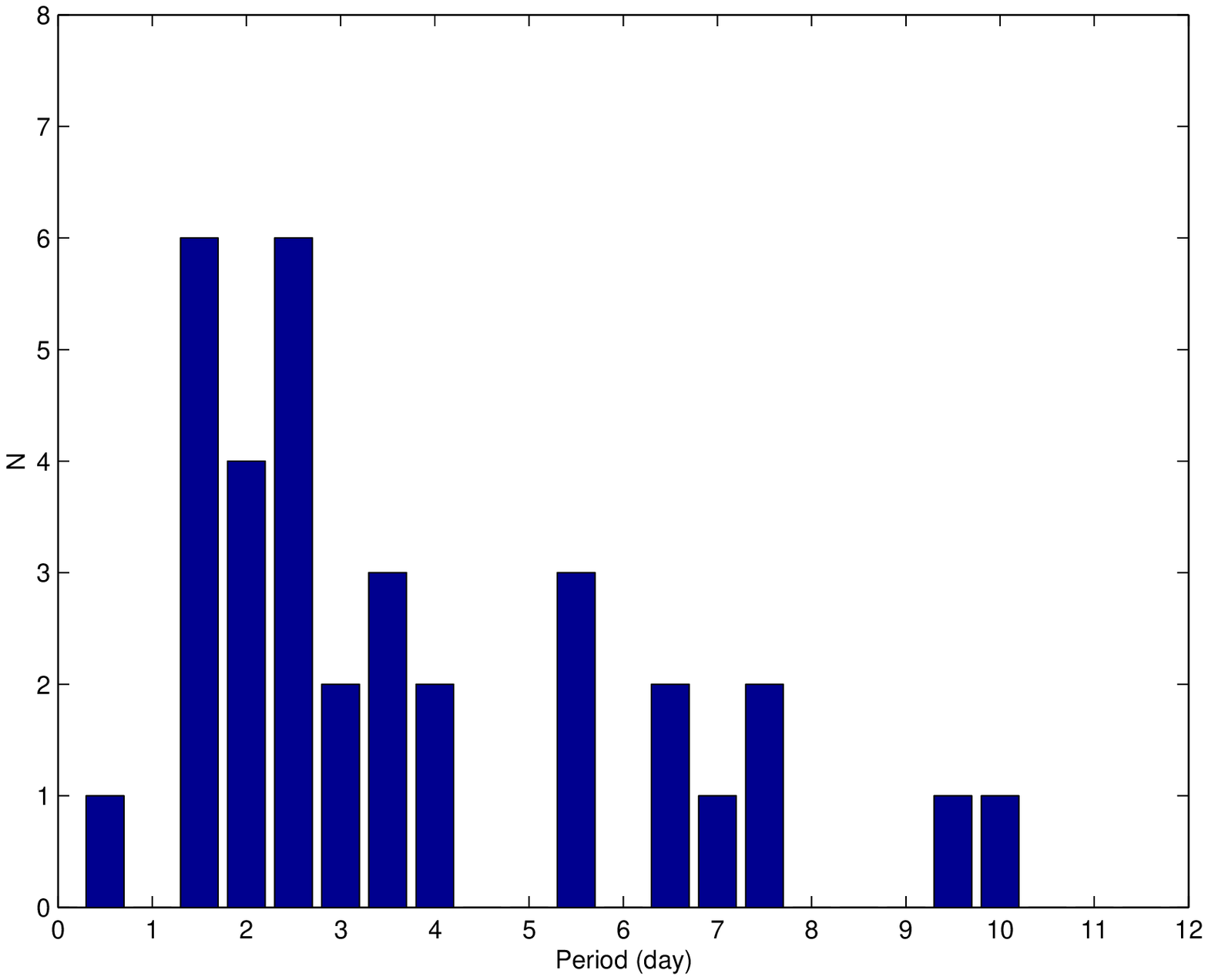}
Fig.2. Stellar rotation period distribution for 36 WTTS with
well-determined periods.
\end{minipage}
\end{figure*}

\begin{figure*}
\includegraphics{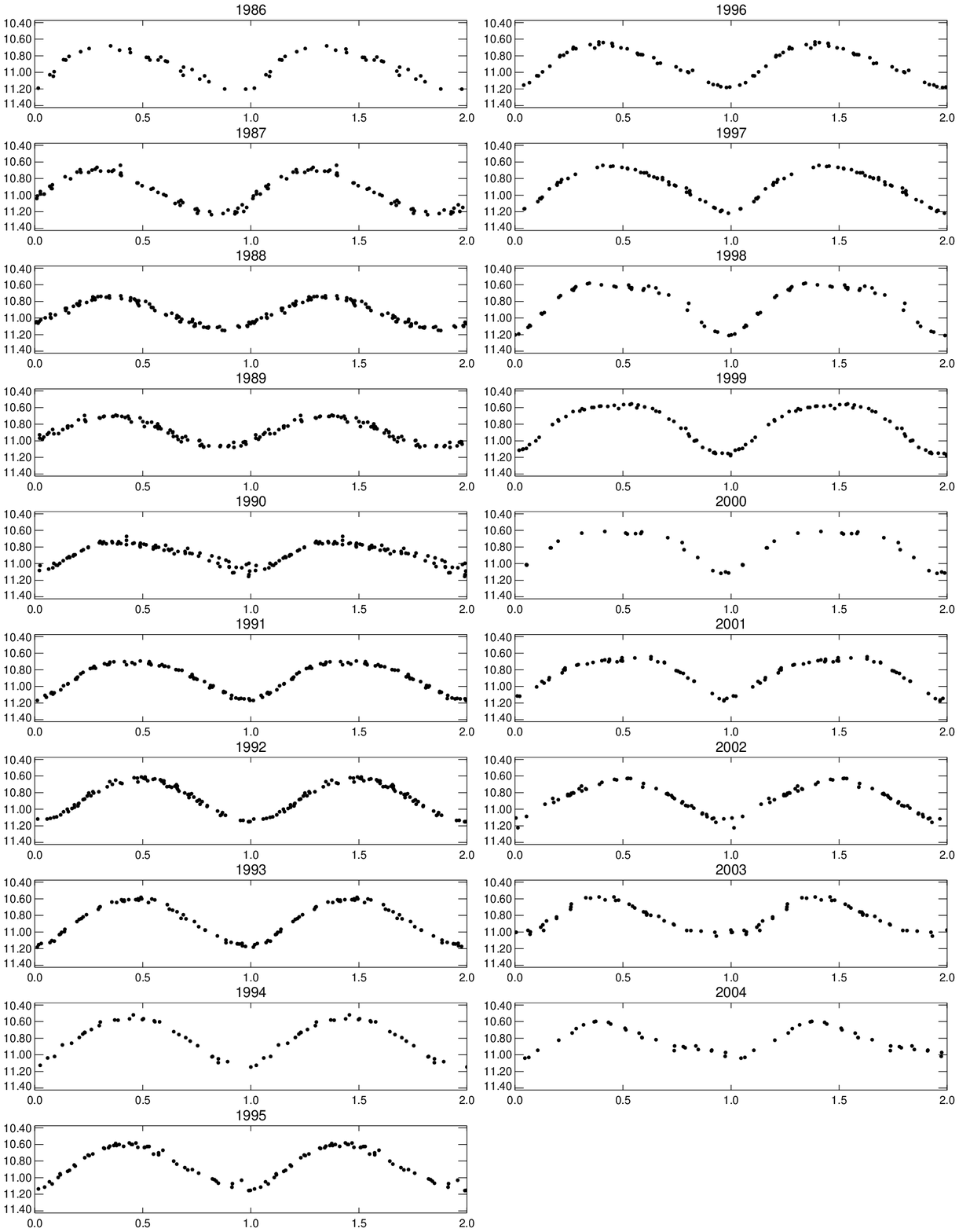}
Fig.3. Phased light curves of V410 Tau for each season. The phased
light curve conserves the phase of light minimum ($\phi_{min}$)
over several years.
\end{figure*}

\begin{figure*}
\includegraphics{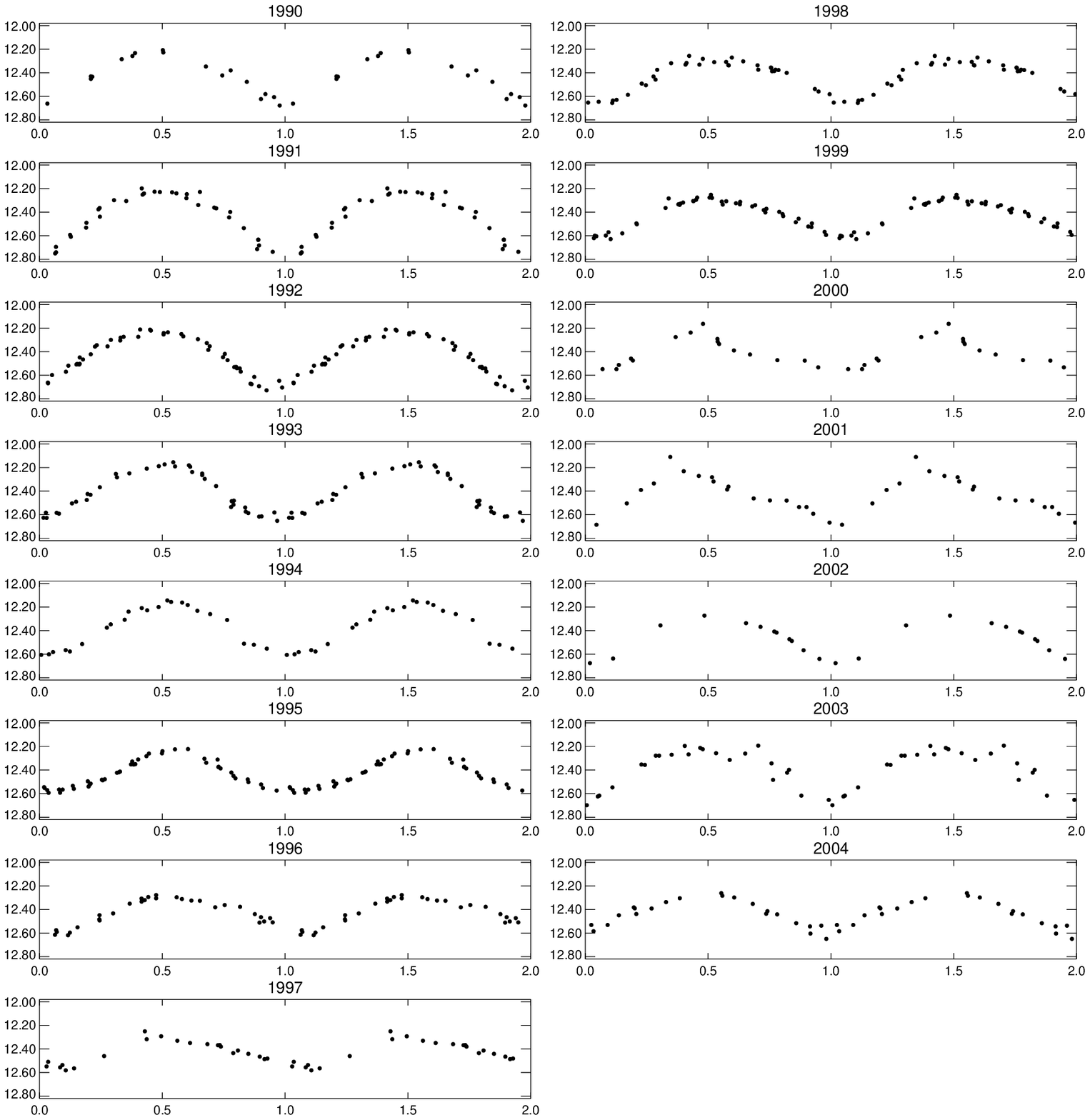}
Fig.4. Phased light curves of LkCa 7. This is another example of
the phenomenon of $\phi_{min}$ stability. At the same time it is
possible to see some changes of amplitude and shape of a light
curve.
\end{figure*}

\begin{figure*}
\centering
\includegraphics{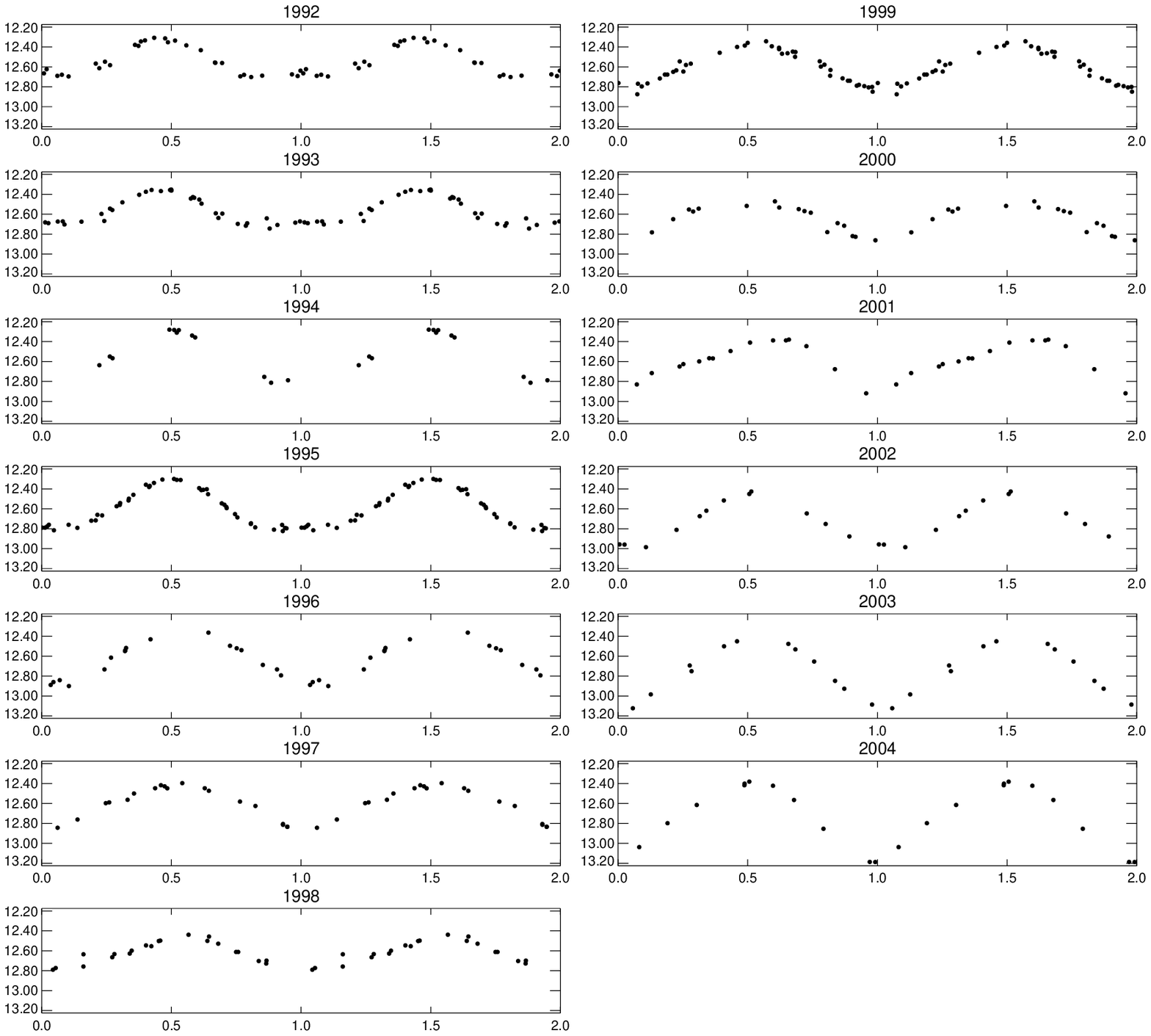}
Fig.5. Phased light curves of LkCa 4. This star showns the highest
amplitude of periodic light variation in our sample ($0.79^m$).
\end{figure*}

\begin{figure*}
\includegraphics{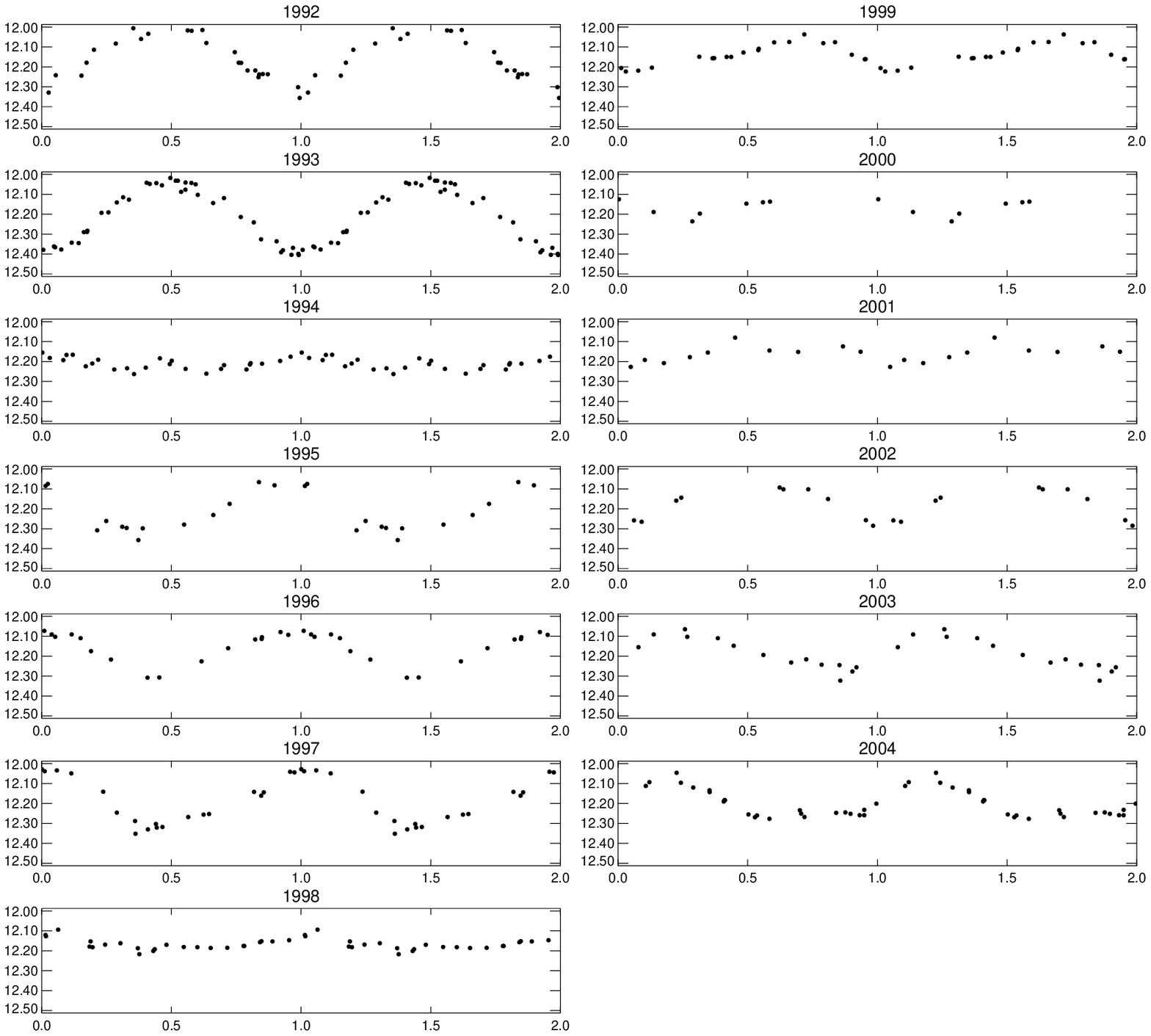}
Fig.6. Phased light curves of TAP 41. The amplitude of the
periodic light variations changes appreciably (by a few tenths of
a magnitude) from season to season.
\end{figure*}

\begin{figure*}
\includegraphics{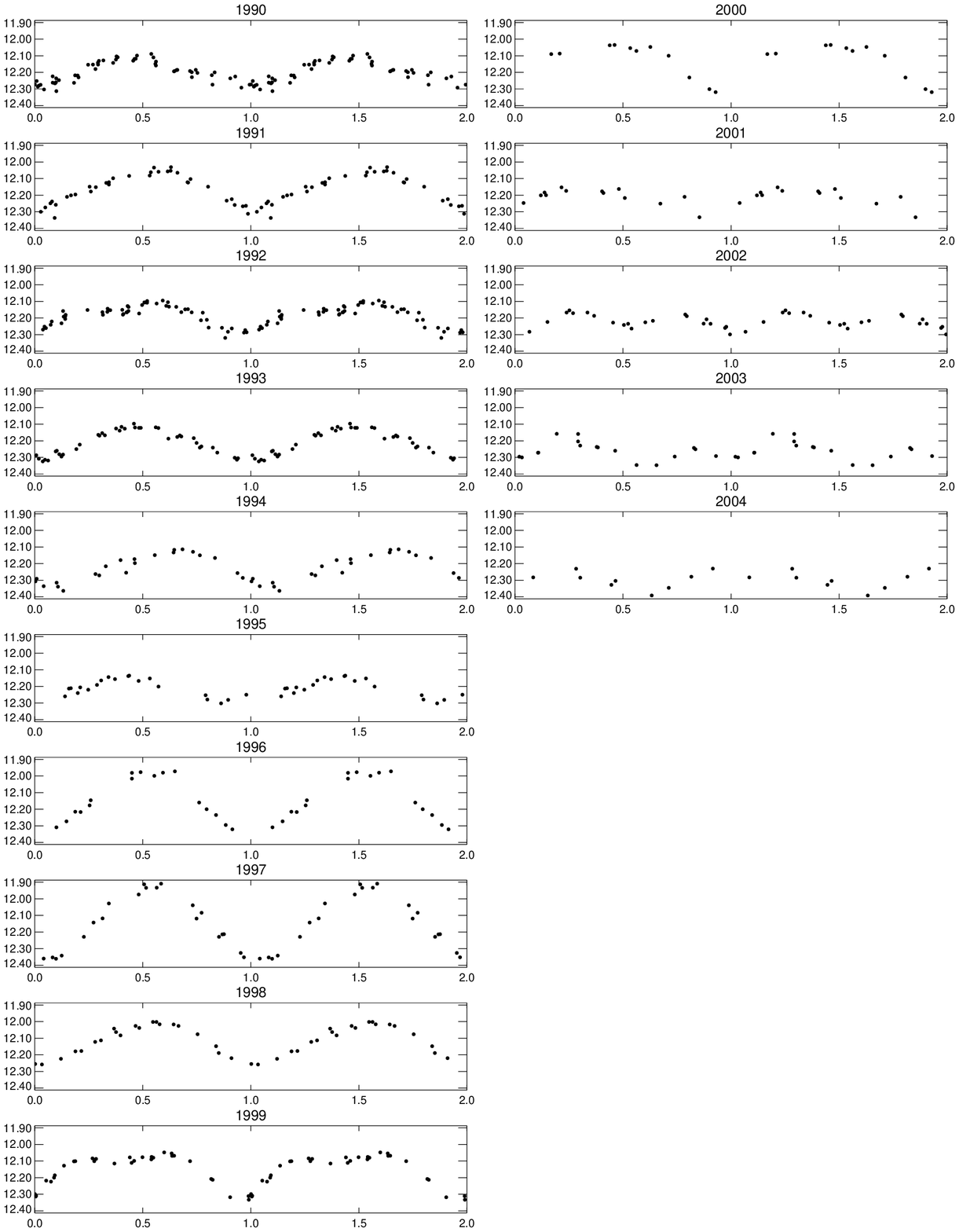}
Fig.7. Phased light curves of V830 Tau. During 2002-2003 the
phased light curve had a complicated shape consisting of two
maxima and two minima during each cycle.
\end{figure*}

\begin{figure*}
\begin{minipage}[c]{8.7cm}
\includegraphics[width=8.5cm]{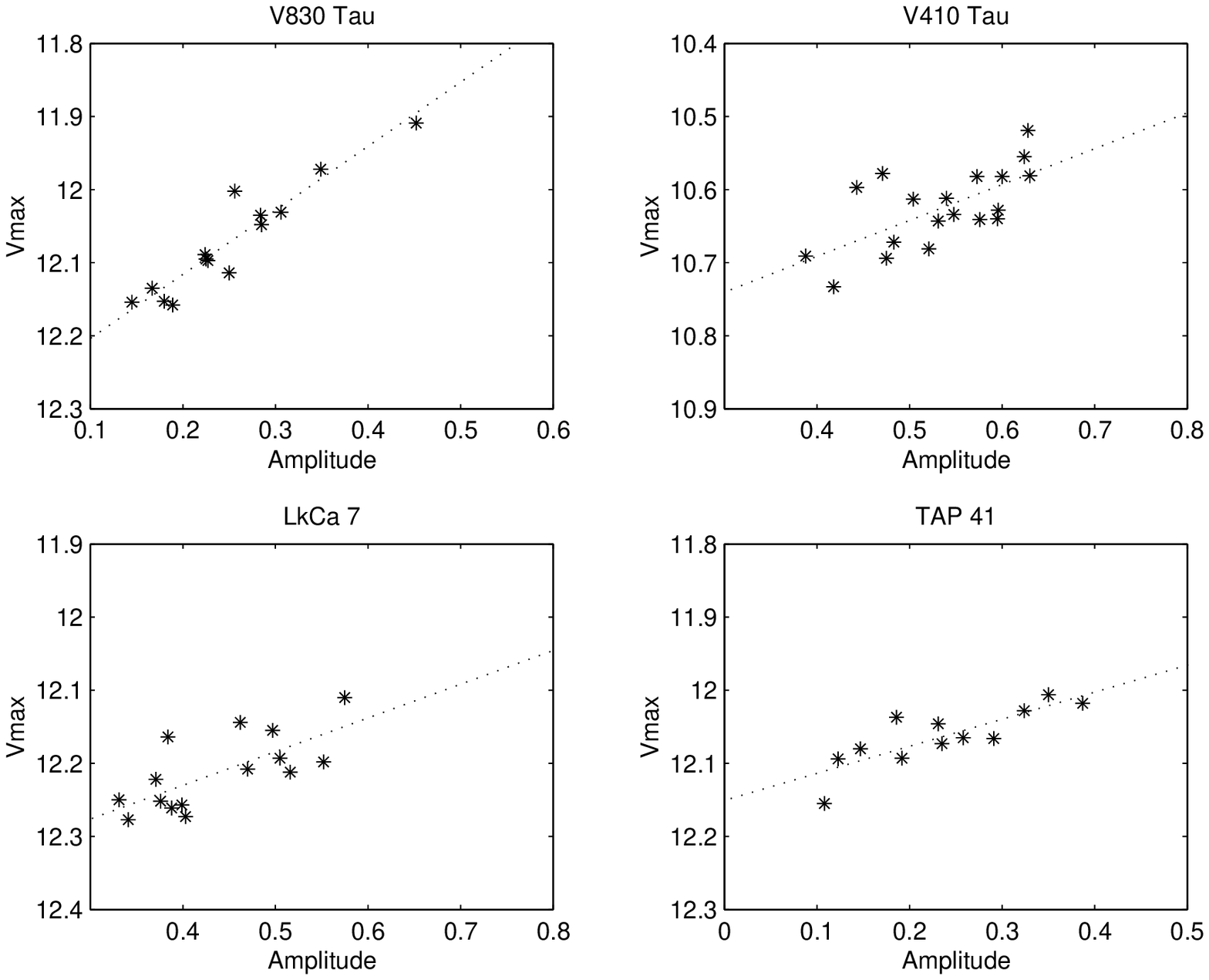}
Fig.8. Illustrations of the linear relation found for some WTTS
between the level of maximum brightness in a season and the
amplitude of the variability during the same season.
\end{minipage}
\hfill
\begin{minipage}[c]{8.7cm}
\includegraphics[width=8.5cm]{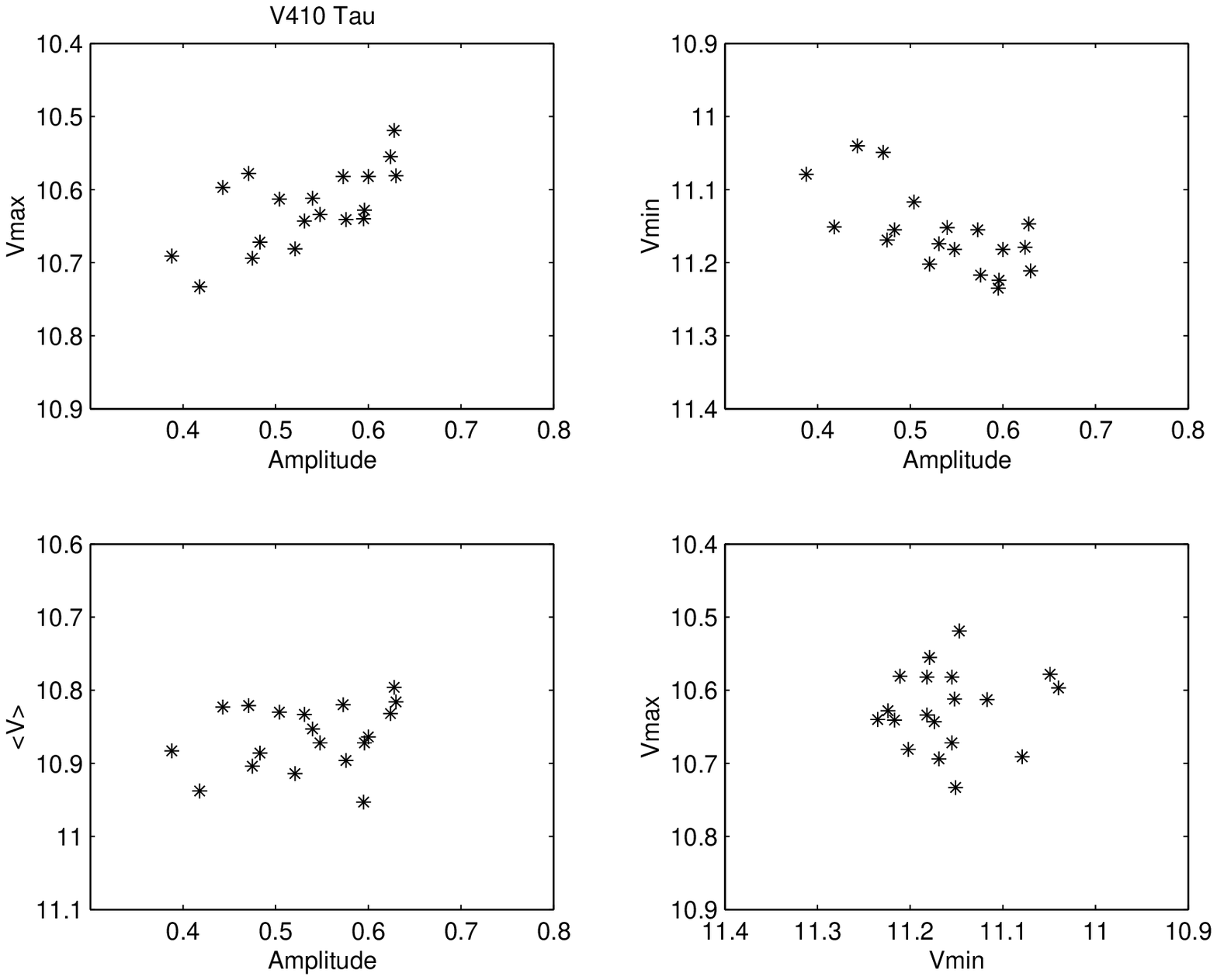}
Fig.9. Dependencies between amplitude, average level of light,
Vmin and Vmax for V410 Tau in different seasons.
\end{minipage}
\end{figure*}

\begin{figure*}
\begin{minipage}[c]{8.7cm}
\includegraphics[width=8.5cm]{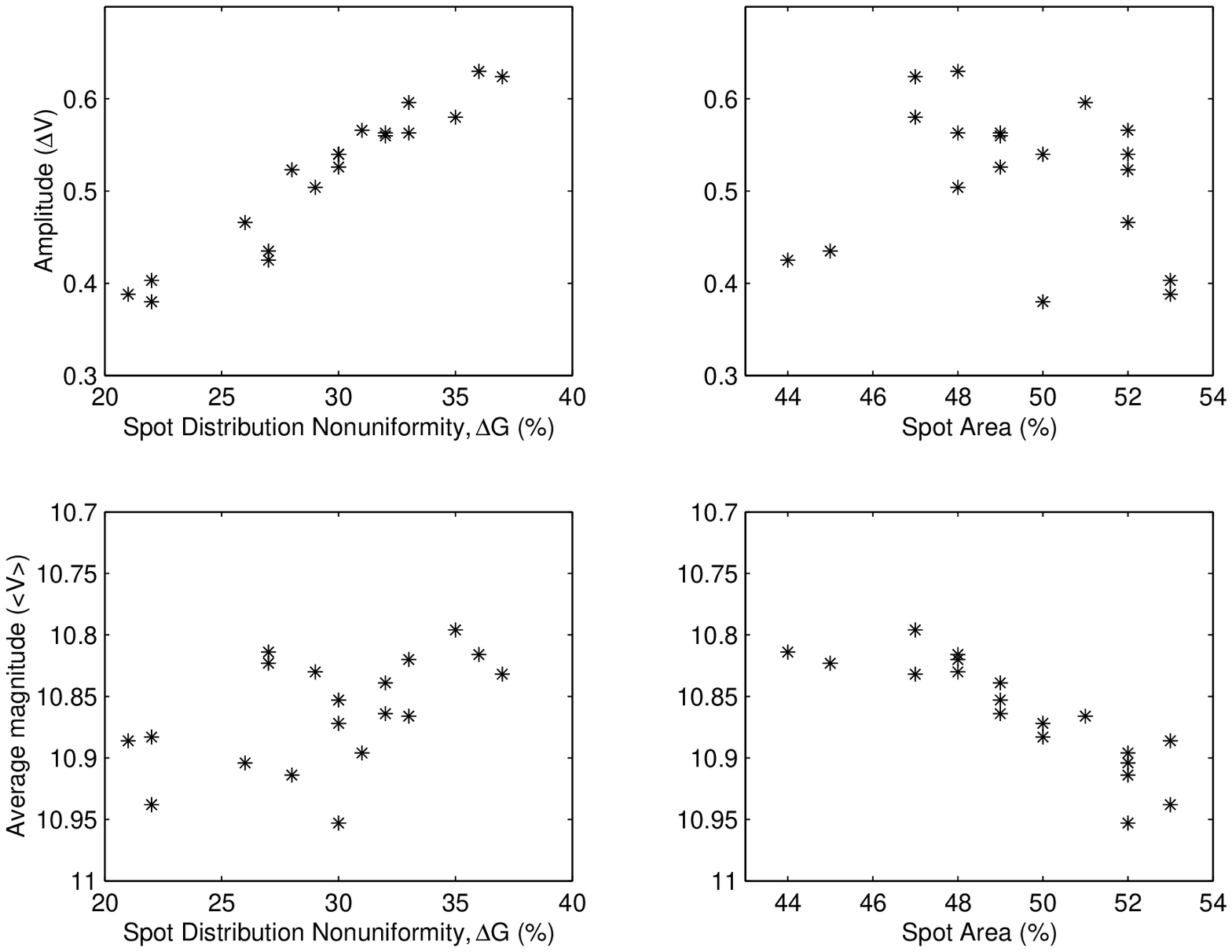}
Fig.10. Plots of the amplitude (top) and the average level of
brightness (bottom) of V410 Tau against the non-uniformity in the
spot distribution (left) and the total spot area (right).
\end{minipage}
\hfill
\begin{minipage}[c]{8.7cm}
\includegraphics[width=8.5cm]{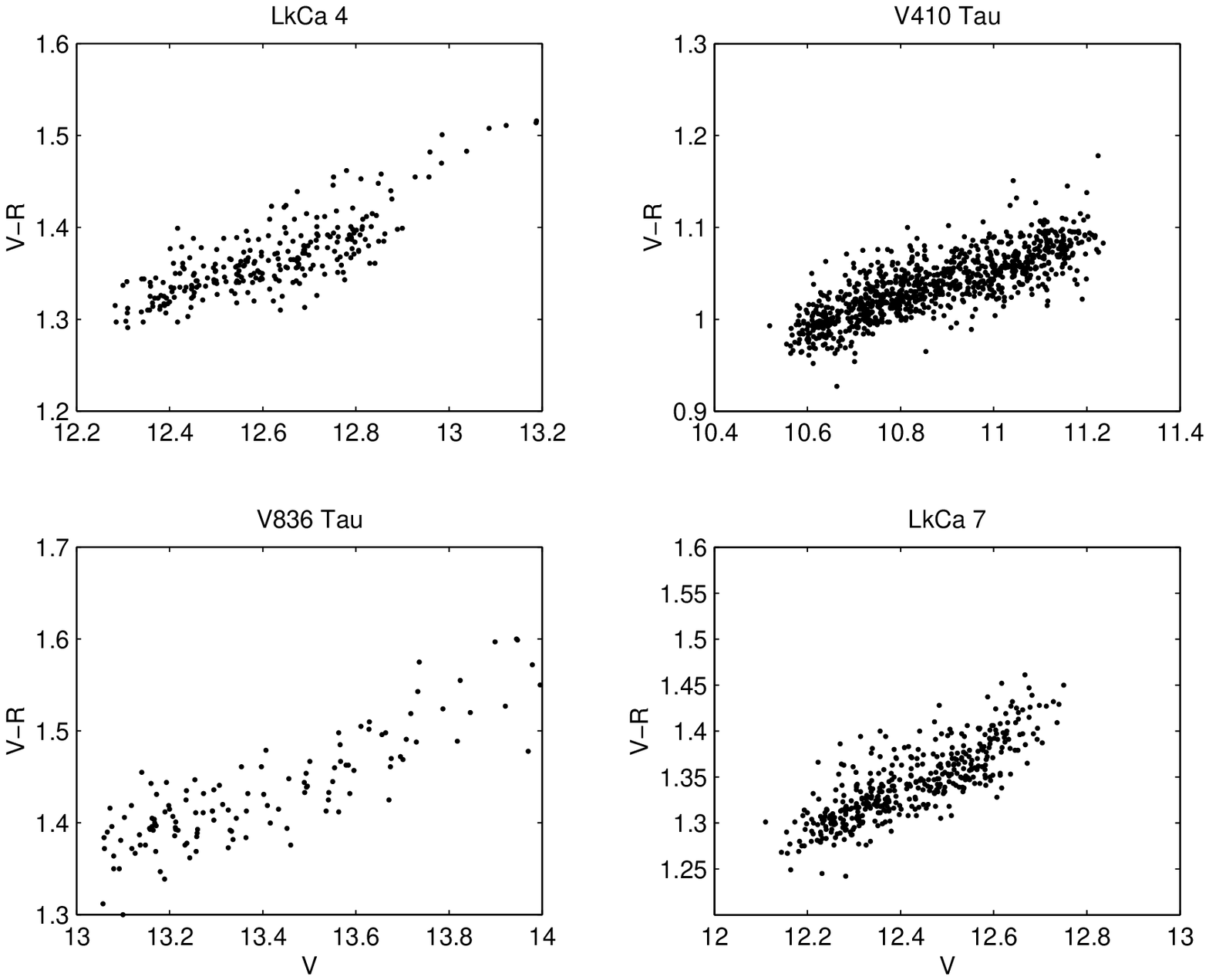}
Fig.11. Illustrations of the linear relations found for most
objects between $V-R$ colour and $V$ magnitude. Colour slopes and
correlation factors are listed in Table 4.
\end{minipage}
\end{figure*}

\begin{figure*}
\centering
\includegraphics{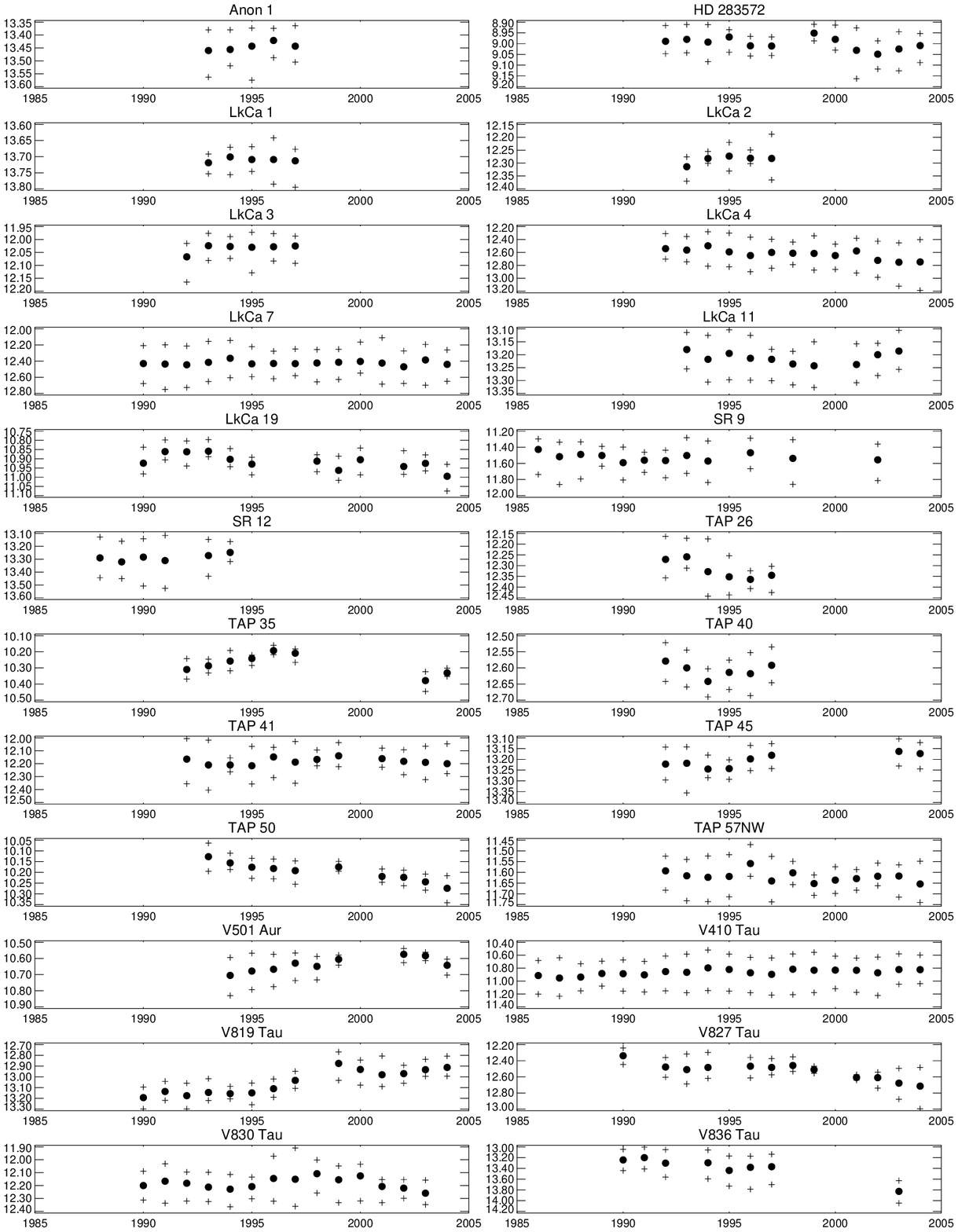}
\flushleft {Fig.12a. Long term V-band light curves of 33 WTTS
observed at Mt. Maidanak. The mean brightness level is shown by a
filled circle for each observing season and the minimum and
maximum brightness values by crosses.}
\end{figure*}


\begin{figure*}
\includegraphics{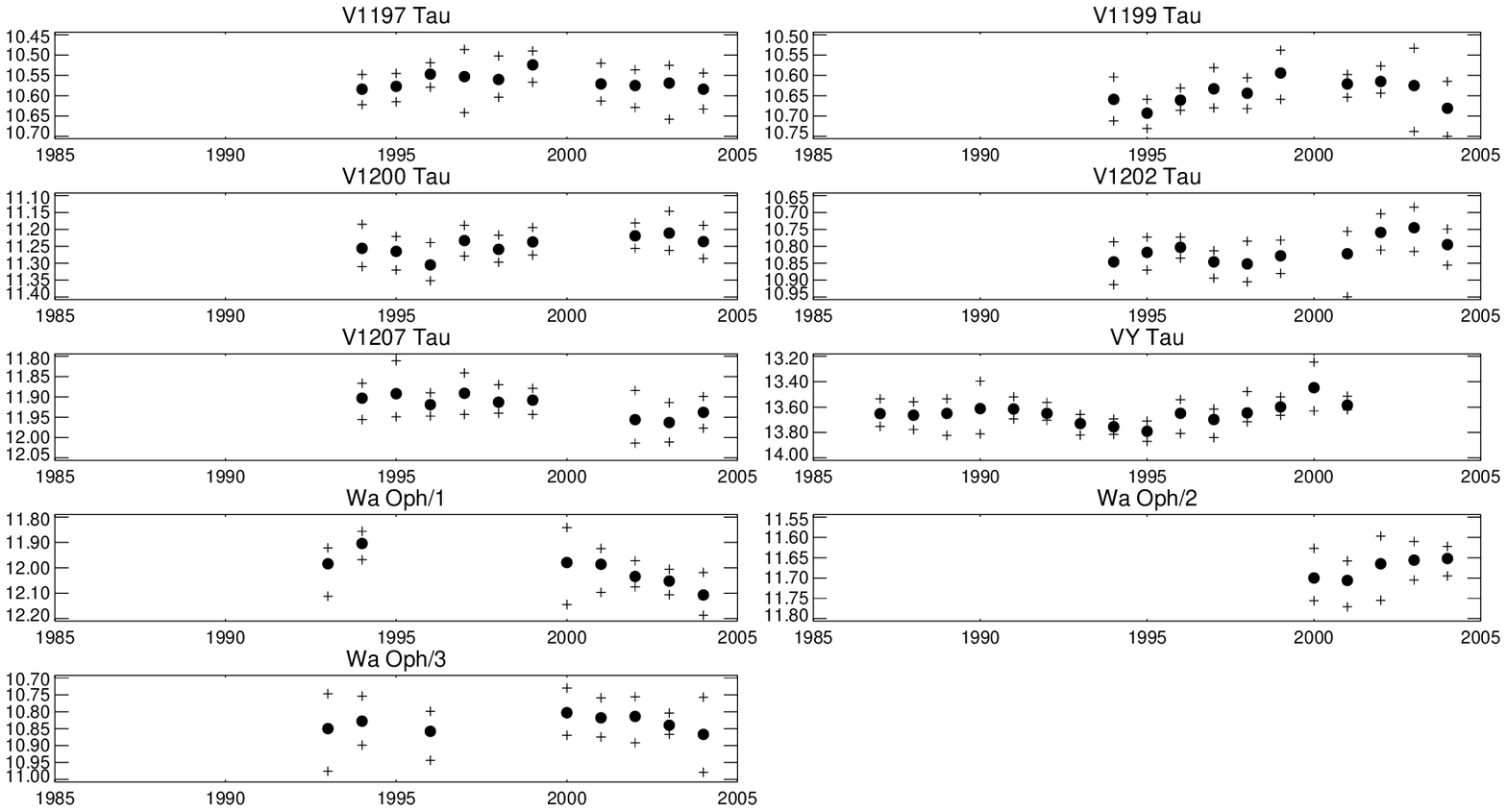}
\flushleft {Fig.12b. (Continue of Fig. 12a.)}
\end{figure*}

\clearpage
\newpage

\begin{figure*}
\begin{minipage}[c]{8.7cm}
\includegraphics[width=8.5cm]{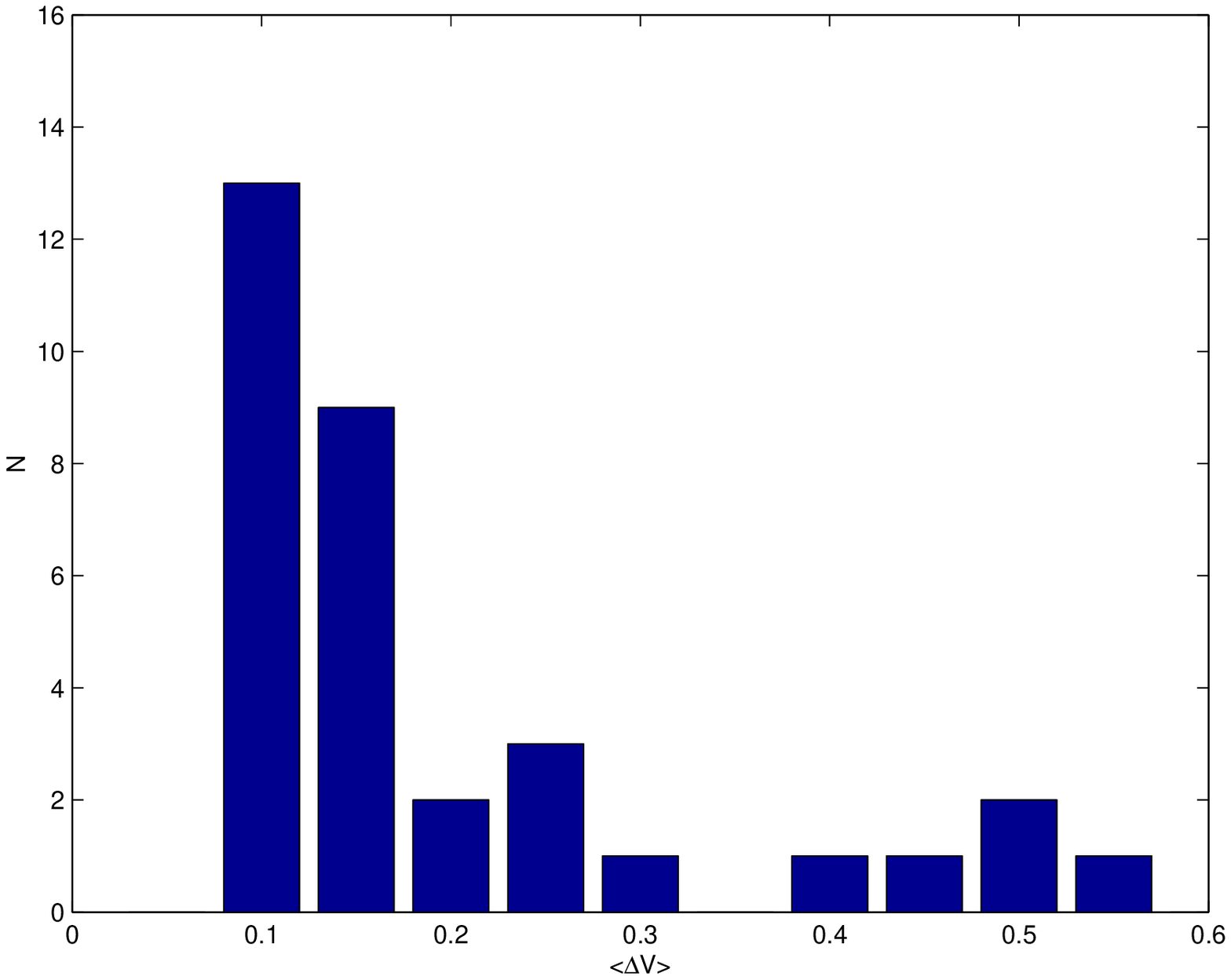}
Fig.13. Average photometric
  amplitude in the $V$-band for the 33 WTTS of our sample
  ($\overline{\Delta V}$).
\end{minipage}
\hfill
\begin{minipage}[c]{8.6cm}
\includegraphics[width=8.4cm]{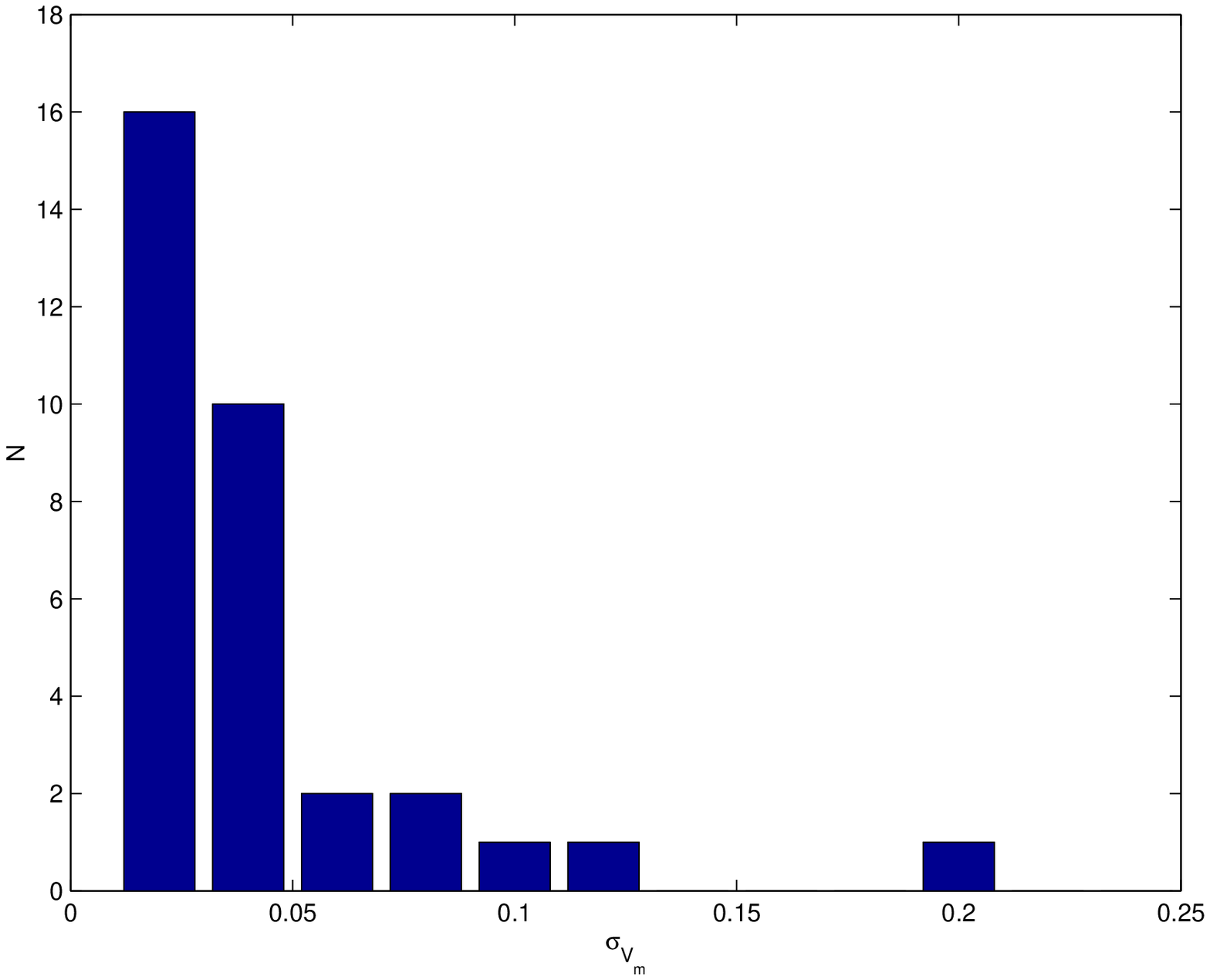}
Fig.14. Standard deviation of the
  seasonal mean brightness level ($\sigma_{V_m}$).
\end{minipage}
\end{figure*}

\begin{figure*}
\begin{minipage}[c]{8.7cm}
\includegraphics[width=8.5cm]{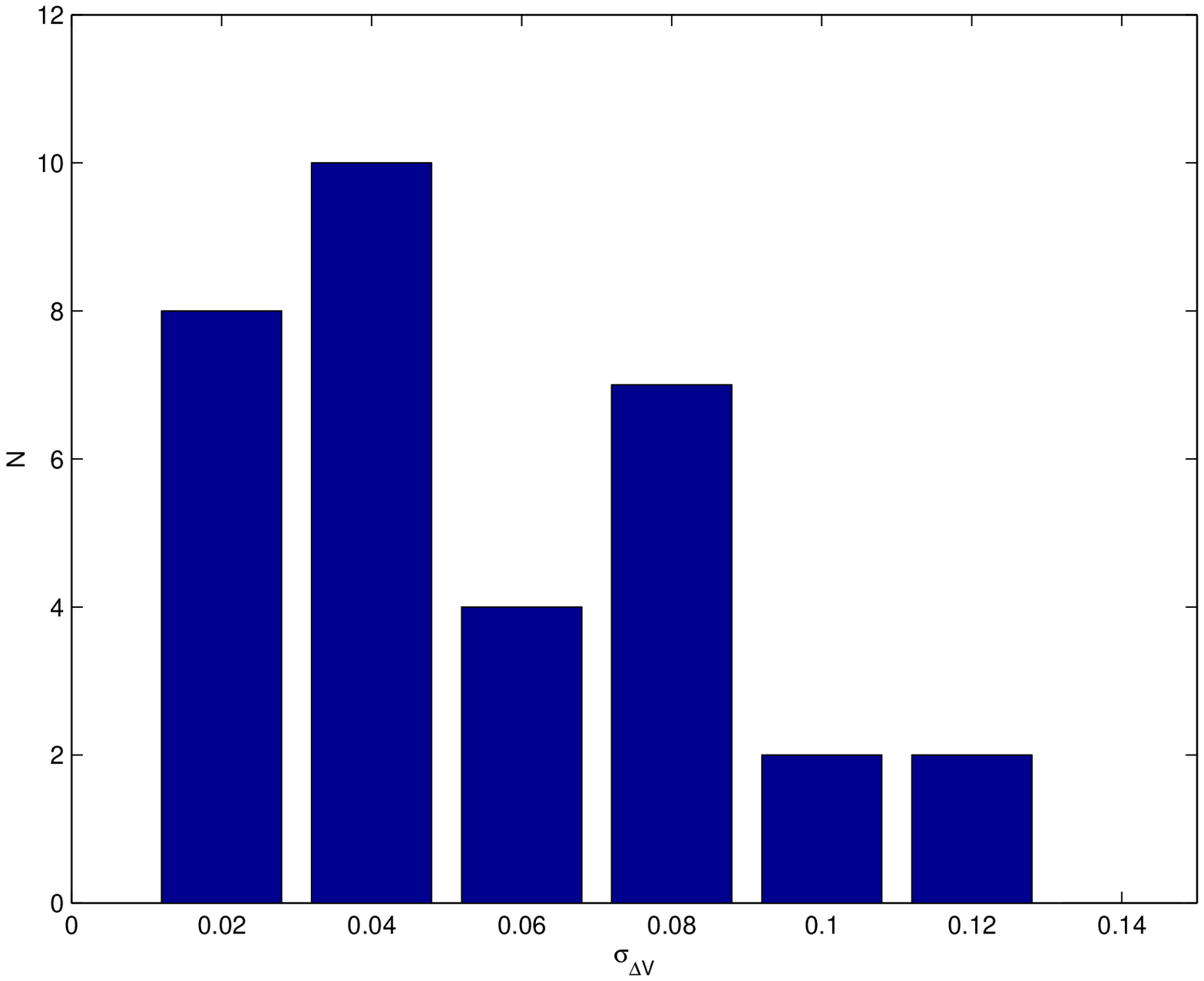}
Fig.15. Standard deviation of the seasonal photometric amplitude
($\sigma_{\Delta V}$).
\end{minipage}
\hfill
\begin{minipage}[c]{8.7cm}
\includegraphics[width=8.5cm]{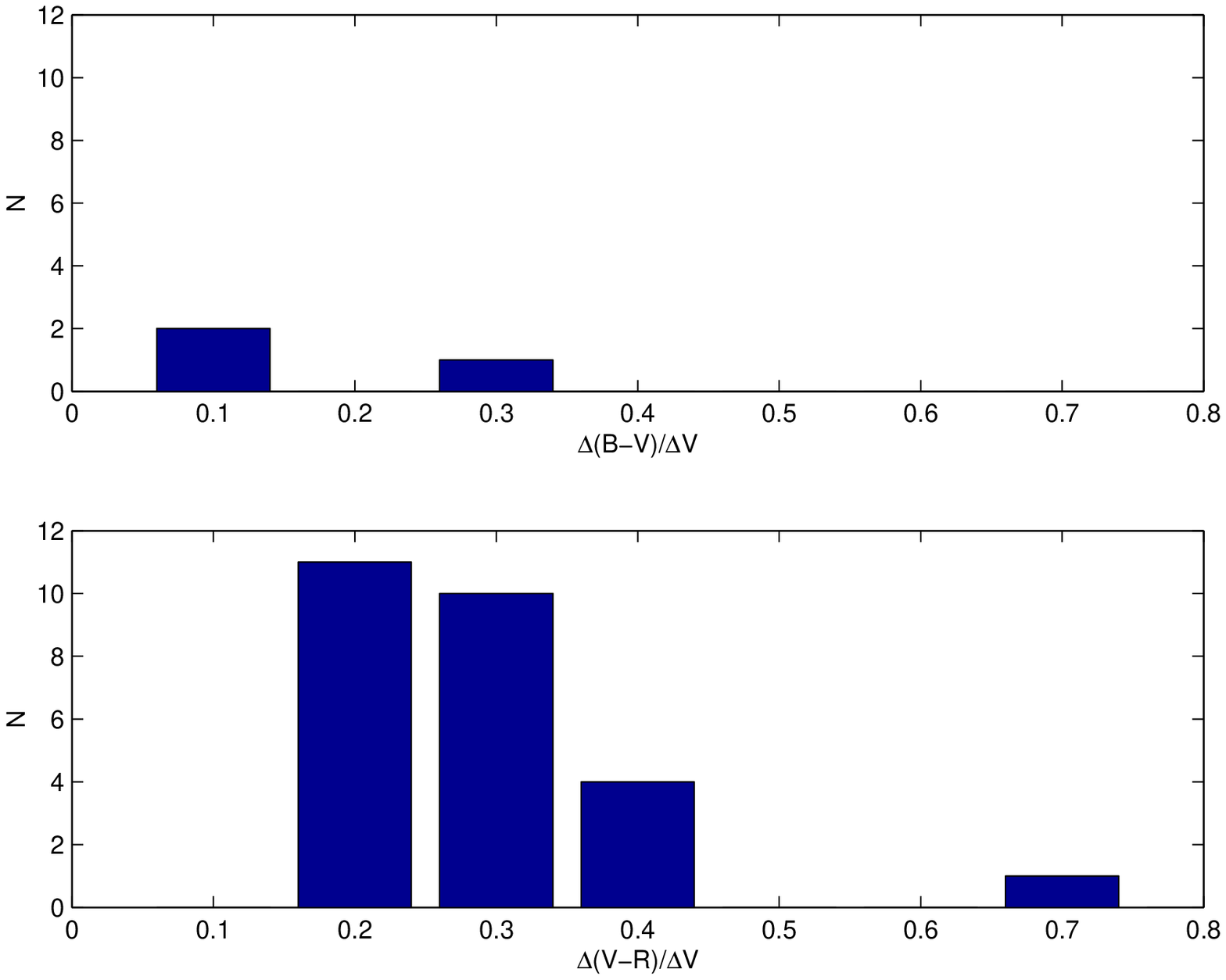}
Fig.16. Least-square color slopes. Top panel : $\frac{\Delta
(B-V)}{\Delta V}$; lower panel: $\frac{\Delta (V-R)}{\Delta V}$.
\end{minipage}
\end{figure*}

\begin{figure*}
\begin{minipage}[t]{8.7cm}
\includegraphics[width=8.5cm]{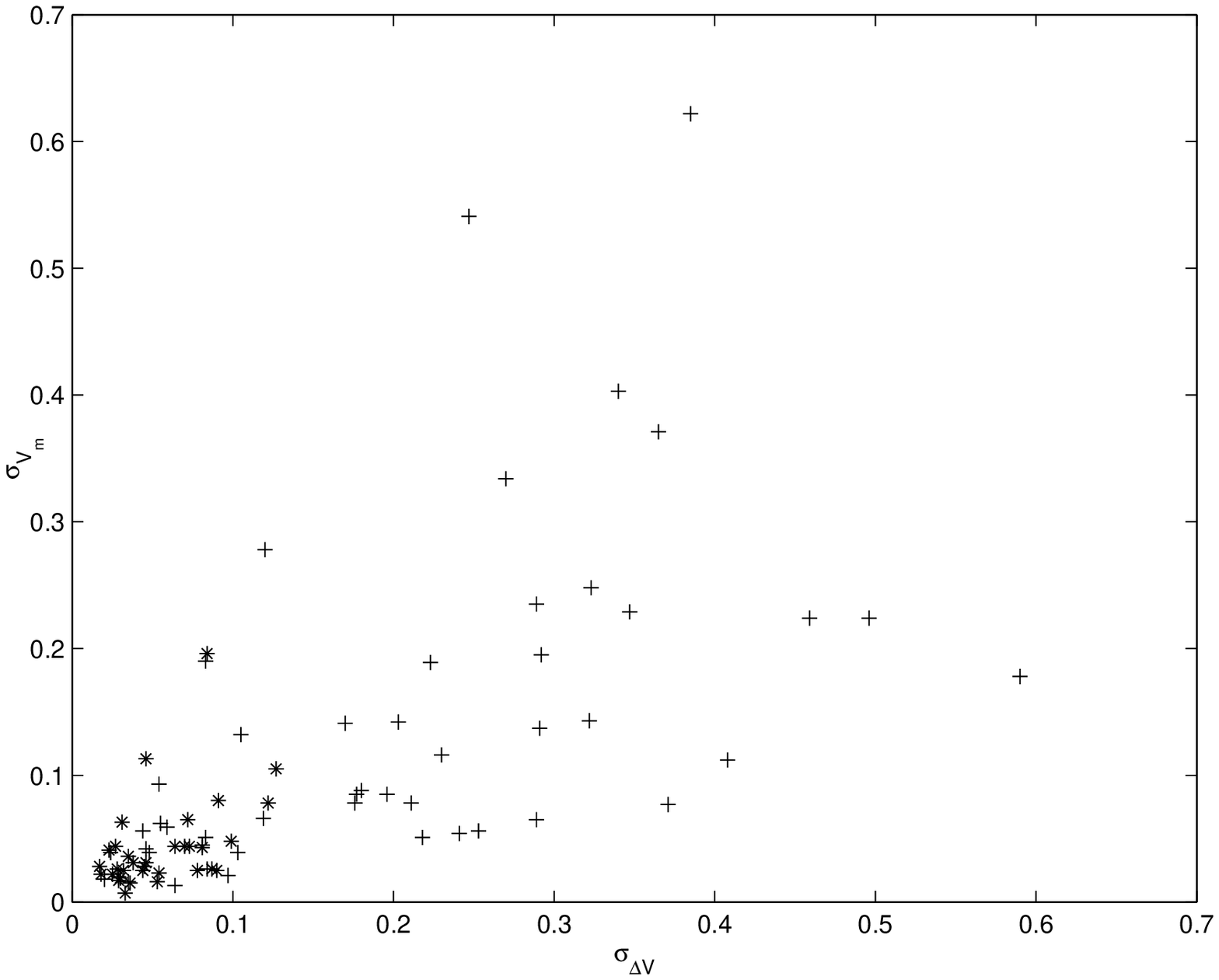}
Fig.17. Variability of the mean brightness level versus the
variability of the photometric amplitude. 33 WTT stars are denoted
by asterisks and 49 CTTs are denoted by crosses.
\end{minipage}
\hfill
\begin{minipage}[t]{8.7cm}
\includegraphics[width=8.5cm]{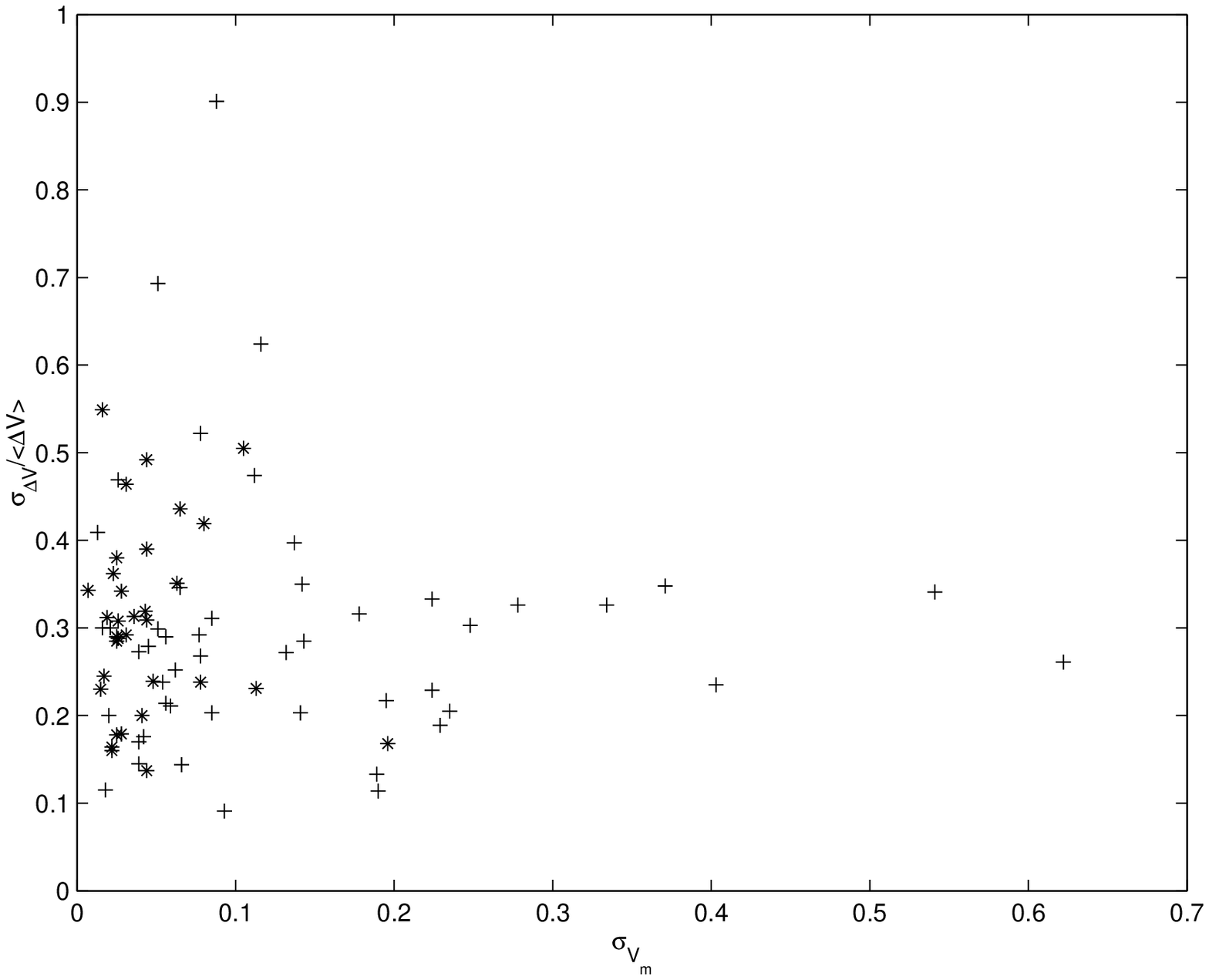}
Fig.18. The relative variation of amplitude versus the variation
of the mean brightness level.
\end{minipage}
\end{figure*}

\begin{figure*}
\begin{minipage}[t]{8.7cm}
\includegraphics[width=8.5cm]{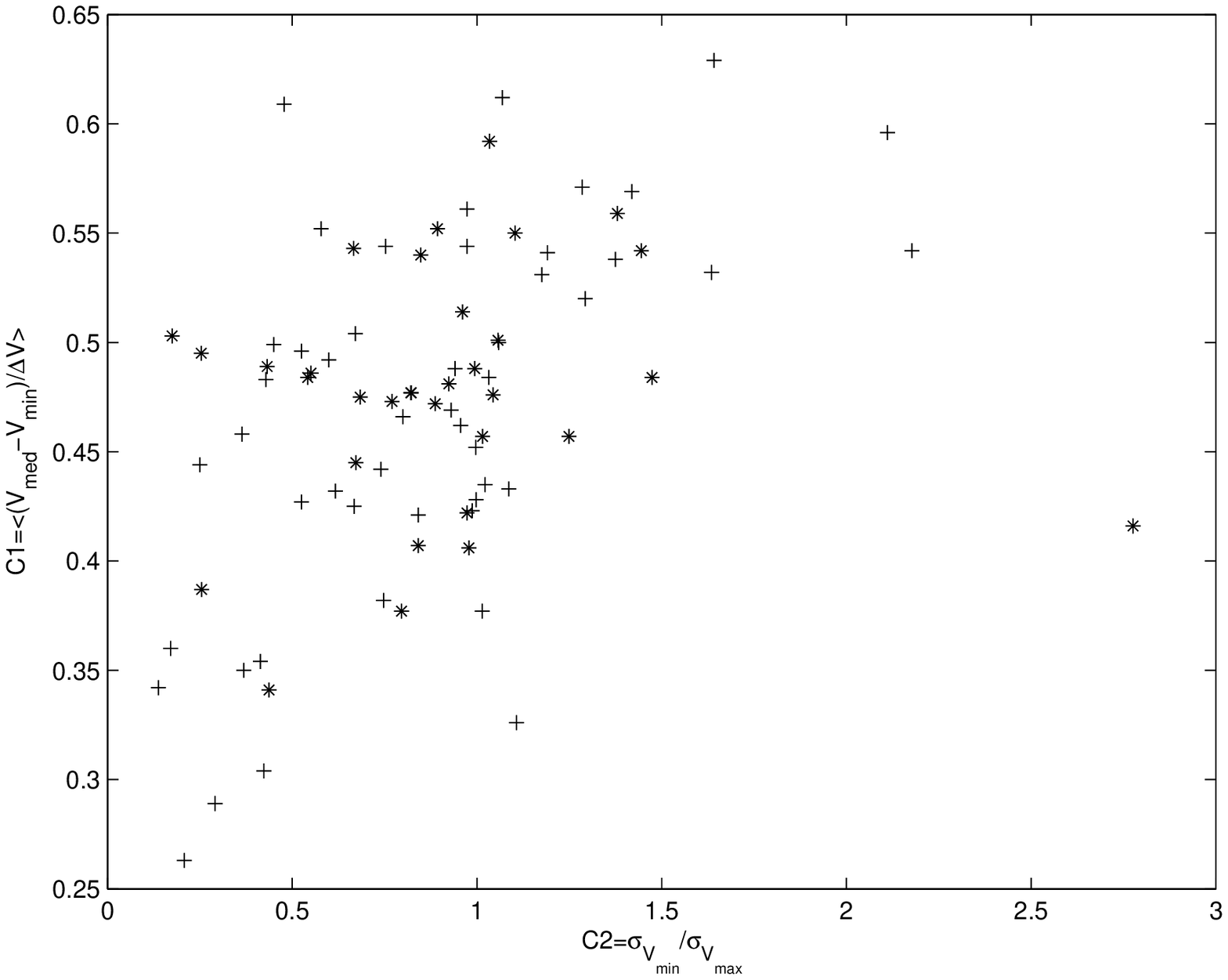}
Fig.19. The star's preferred brightness state (as measured by
 C1=${<\frac{V_{med} - V_{min}}{\Delta V}>}$) is plotted against the relative
 variations of its extreme brightness states (as measured by
 C2=$\sigma_{V_{min}}/\sigma_{V_{max}}$). See text.
\end{minipage}
\hfill

\end{figure*}

\end{document}